\documentclass[useAMS,usenatbib]{mn2e}

\def\jnl@style#1{{\rmfamily#1}}%
\def\jref@jnl#1{{\jnl@style#1}}%

\newcommand\aj{\jref@jnl{AJ}}%
\newcommand\araa{\jref@jnl{ARA\&A}}%
\newcommand\apj{\jref@jnl{ApJ}}%
\newcommand\apjl{\jref@jnl{ApJ}}%
\newcommand\apjs{\jref@jnl{ApJS}}%
\newcommand\ao{\jref@jnl{Appl.~Opt.}}%
\newcommand\apss{\jref@jnl{Ap\&SS}}%
\newcommand\aap{\jref@jnl{A\&A}}%
\newcommand\aapr{\jref@jnl{A\&A~Rev.}}%
\newcommand\aaps{\jref@jnl{A\&AS}}%
\newcommand\azh{\jref@jnl{AZh}}%
\newcommand\baas{\jref@jnl{BAAS}}%
\newcommand\jrasc{\jref@jnl{JRASC}}%
\newcommand\memras{\jref@jnl{MmRAS}}%
\newcommand\mnras{\jref@jnl{MNRAS}}%
\newcommand\na{\jref@jnl{New~Astron.}}
\newcommand\pra{\jref@jnl{Phys.~Rev.~A}}%
\newcommand\prb{\jref@jnl{Phys.~Rev.~B}}%
\newcommand\prc{\jref@jnl{Phys.~Rev.~C}}%
\newcommand\prd{\jref@jnl{Phys.~Rev.~D}}%
\newcommand\pre{\jref@jnl{Phys.~Rev.~E}}%
\newcommand\prl{\jref@jnl{Phys.~Rev.~Lett.}}%
\newcommand\pasp{\jref@jnl{PASP}}%
\newcommand\pasj{\jref@jnl{PASJ}}%
\newcommand\qjras{\jref@jnl{QJRAS}}%
\newcommand\skytel{\jref@jnl{S\&T}}%
\newcommand\solphys{\jref@jnl{Sol.~Phys.}}%
\newcommand\sovast{\jref@jnl{Soviet~Ast.}}%
\newcommand\ssr{\jref@jnl{Space~Sci.~Rev.}}%
\newcommand\zap{\jref@jnl{ZAp}}%
\newcommand\nat{\jref@jnl{Nature}}%
\newcommand\iaucirc{\jref@jnl{IAU~Circ.}}%
\newcommand\aplett{\jref@jnl{Astrophys.~Lett.}}%
\newcommand\apspr{\jref@jnl{Astrophys.~Space~Phys.~Res.}}%
\newcommand\bain{\jref@jnl{Bull.~Astron.~Inst.~Netherlands}}%
\newcommand\fcp{\jref@jnl{Fund.~Cosmic~Phys.}}%
\newcommand\gca{\jref@jnl{Geochim.~Cosmochim.~Acta}}%
\newcommand\grl{\jref@jnl{Geophys.~Res.~Lett.}}%
\newcommand\jcp{\jref@jnl{J.~Chem.~Phys.}}%
\newcommand\jgr{\jref@jnl{J.~Geophys.~Res.}}%
\newcommand\jqsrt{\jref@jnl{J.~Quant.~Spec.~Radiat.~Transf.}}%
\newcommand\memsai{\jref@jnl{Mem.~Soc.~Astron.~Italiana}}%
\newcommand\nphysa{\jref@jnl{Nucl.~Phys.~A}}%
\newcommand\physrep{\jref@jnl{Phys.~Rep.}}%
\newcommand\physscr{\jref@jnl{Phys.~Scr}}%
\newcommand\planss{\jref@jnl{Planet.~Space~Sci.}}%
\newcommand\procspie{\jref@jnl{Proc.~SPIE}}%

\usepackage{amssymb, amsmath}
\usepackage{bm}
\usepackage{graphicx}
\usepackage{float}
\usepackage{color}
\usepackage[font=footnotesize]{subfig}
\usepackage{changes}
\usepackage[abs]{overpic}
\usepackage{soul}
\usepackage{hyperref}

\newcommand{\CII}{[C$\,${\sc ii}]}
\newcommand{\CIIzwei}{[$^{12}$C$\,${\sc ii}]}
\newcommand{\CIIdrei}{[$^{13}$C$\,${\sc ii}]}

\newcommand{\Ltot}{$L_{\textrm{tot}}$}

\newcommand{\HI}{H$\,${\sc i}}

\newcommand{\kms}{km~s$^{-1}$}

\newcommand{\cmdrei}{cm$^{-3}$}
\newcommand{\YCII}{$Y_{{\rm CII}}$}

\newcommand{\IXCII}{cm$^{-2}$ (K~km~s$^{-1}$)$^{-1}$}

\newcommand{\Tex}{$T_{\textrm{ex}}$}
\newcommand{\Tkin}{$T_{\textrm{kin}}$}

\newcommand{\RADMC}{{\sc radmc--3d}}
\newcommand{\FLASH}{{\sc flash}}
\newcommand{\AREPO}{{\sc arepo}}

\title[Synthetic \CII\ maps of a molecular cloud]
{Synthetic \CII\  emission maps of a simulated molecular cloud  in formation}
\author[A. Franeck et al.]{A. Franeck$^{1}$\thanks{E-mail:
franeck@ph1.uni-koeln.de (AF)}, S. Walch$^{1}$, D. Seifried$^{1}$, S.D. Clarke$^{1}$, V. Ossenkopf-Okada$^{1}$
\newauthor 
S.C.O. Glover$^{2}$, R.S. Klessen$^{2, 3}$, P. Girichidis$^{4}$, T. Naab$^{5}$, R. W\"unsch$^{6}$,
\newauthor
P.C. Clark$^{7}$, E. Pellegrini$^{2}$, T. Peters$^{5}$ \\
$^{1}$I. Physikalisches Institut, Universit\"at zu K\"oln, Z\"ulpicher Stra\ss e 77, D-50937 K\"oln, Germany\\
$^{2}$Universit\"at Heidelberg, Zentrum f\"ur Astronomie, Institut f\"ur Theoretische Astrophysik, Albert-Ueberle-Stra\ss e 2, D-69120 Heidelberg, Germany\\
$^{3}$Universit\"at Heidelberg, Institut f\"ur wissenschaftliches Rechnen (IWR), Im Neuenheimer Feld 205, 69120 Heidelberg\\
$^{4}$Leibniz-Institut f\"ur Astrophysik Potsdam (AIP), An der Sternwarte 16, D-14482 Potsdam, Germany \\
$^{5}$Max-Planck-Institut f\"ur Astrophysik, Karl-Schwarzschild-Stra\ss e 1, D-85748 Garching, Germany\\
$^{6}$Astronomical Institute, Czech Academy of Sciences, Bocni II
1401, CZ-141 00 Prague, Czech Republic\\
$^{7}$School of Physics \& Astronomy, Cardiff University, 5 The Parade, Cardiff CF24 3AA, Wales, UK}
\begin{document}

\date{Accepted 2018 September 11; Revised 2018 August 25; Received: 2018 May 30}

\pagerange{\pageref{firstpage}--\pageref{lastpage}} \pubyear{2018}

\maketitle

\label{firstpage}

\begin{abstract}
The C$^{+}$ ion is an important coolant of interstellar gas, and so the \CII\ fine structure line is frequently observed in the interstellar medium.
However, the physical and chemical properties of the \CII-emitting gas are still unclear. We carry out non-LTE  radiative transfer simulations with \RADMC\ to study the \CII\ line emission from a young, turbulent molecular cloud before the onset of star formation, using data from the SILCC-Zoom project. 
The \CII\ emission is optically thick over 40\% of the observable area with $I_{[\textrm{CII}]} > 0.5$~K~\kms. To determine the physical properties of the \CII\ emitting gas, we treat the \CII\ emission as optically thin. 
We find that the \CII\ emission originates primarily from cold, moderate density gas  ($40 \lesssim T \lesssim 65$~K and $50 \lesssim n \lesssim 440$~cm$^{-3}$), composed mainly of atomic hydrogen and with an effective visual extinction between $\sim 0.50$ and $\sim 0.91$. Gas dominated by molecular hydrogen contributes only $\lesssim$20\% of the total \CII\ line emission. Thus, \CII\ is not a good tracer for CO-dark H$_2$ at this early phase in the cloud's lifetime. 
We also find that the total gas, H and C$^+$ column densities are all correlated with the integrated \CII\ line emission, with power law slopes ranging from 0.5 to 0.7. 
Further, the median ratio between the total column density and the \CII\ line emission is $Y_{{\rm CII}}\approx 1.1 \times 10^{21}$~\IXCII, and \YCII\ scales with $I_{[\textrm{CII}]}^{-0.3}$. We expect \YCII\ to change in environments with a lower or higher radiation field than simulated here.

\end{abstract}


\section{Introduction}
\label{sec.Intro}

The \CII\ line of ionized carbon is one of the dominant coolants in the interstellar medium \citep[ISM;][]{Tielens.Hollenbach.1985, Klessen.Glover.2016}. 
It originates from the fine structure transition \mbox{$^2P_{3/2} \rightarrow$ $^2P_{1/2}$} of singly ionized carbon (C$^+$). The transition occurs at a wavelength of \mbox{$\lambda_{[\textrm{CII}]} = 157.741$~$\mu$m} (\mbox{$\nu_{[\textrm{CII}]} = 1900.537$~GHz}). 
Only 11.26~eV are needed to ionize carbon atoms to C$^+$, and therefore this ion can coexist with neutral hydrogen, which is ionized by photons with energies equal to or more than 13.6~eV. 
On the other hand, 24.4~eV is necessary to bring C$^+$ into a higher ionization state \citep{Goldsmith.etal.2012, Pineda.etal.2013, Kapala.etal.2015}, and so C$^+$ is still present in the warm, moderately ionized medium. 
It is therefore not trivial to observationally disentangle from which phase of the ISM the \CII\ line emission signal originates. \\

\CII\ line emission accounts for $0.1$--$1$\% of the total infrared emission from the Milky Way \citep{Stacey.etal.1991} and is produced in photon dominated regions \citep[PDRs, e.g.][]{Ossenkopf.etal.2013}, in shock fronts \citep{Appleton.etal.2013, Lesaffre.etal.2013}, but also in forming molecular clouds \citep[e.g.][]{Beuther.etal.2014}. 
Within the GOT~C+ survey\footnote{GOT~C+: Galactic Observations of Terahertz C+ project, mapping the \CII\ line emission in the Milky Way with \textit{Herschel}/HIFI}, the \CII\ line emission was studied in the Milky Way along 452 lines of sight, distributed in an approximately volume-weighted fashion in the Galactic plane \citep{Pineda.etal.2013}. Analysing these data and correlating the \CII\ line emission with the emission of CO and \HI, \cite{Velusamy.Langer.2014} found that $\sim$62\% of the \CII\ line emission is associated with H$_2$ gas, $\sim$18\% with \HI\ gas and $\sim$21\% with the warm interstellar medium (WIM). However, according to \cite{Pineda.etal.2013, Pineda.etal.2014} between $\sim$30\% and $\sim$47\% of the \CII\ line emission originates from PDRs. 
Therefore, these values vary depending on the regions analysed. 
Since the \CII\ line emission is observed in connection with molecular clouds, there have been attempts to use it as a tracer for star formation \citep[e.g.][]{DeLooze.etal.2011, DeLooze.etal.2014, Pineda.etal.2014, Herrera-Camus.etal.2015, Kapala.etal.2015} as well as for CO-dark H$_2$ gas \citep{Wolfire.etal.2010, Langer.etal.II.2014}. 
However, to judge its suitability as a tracer, it is essential to understand from which phase of the ISM the \CII\ line emission originates. 
In this work we study this question with numerical simulations of a forming molecular cloud, before the onset of star formation. \\

Synthetic \CII\ line observations of the ISM were carried out in earlier studies, e.g.\ by \cite{Accurso.etal.2017}. 
They run a suite of 1D models, coupling  \textsc{starburst99} \citep{Leitherer.etal.1999, Leitherer.etal.2010, Vazquez.Leitherer.2005, Conroy.2013}, \textsc{mocassin} \citep{Ercolano.etal.2003} and \textsc{3d-pdr} \citep{Bisbas.etal.2012} to find the \CII\ line emission as a function of e.g.\ the specific star formation rate, the gas phase metallicity, the electron number density, and the dust mass fraction. For a galaxy in the local neighbourhood they predict 60--80\% of the \CII\ line emission to originate from regions dominated by molecular gas. 
Synthetic \CII\ line emission studies on the scales of molecular clouds have been carried out by \cite{Bisbas.etal.2017} and \cite{Bertram.etal.2016}. The work of \cite{Bisbas.etal.2017} investigates whether \CII\ line emission can be used as a tracer for giant molecular cloud collisions. 
\cite{Bertram.etal.2016} study a molecular cloud in a Galactic Center like environment. They compare the difference in the evolution of the molecular clouds when the virial parameter is changed by varying the amount of kinetic energy in the simulation. 
Overall, they find that atomic tracers like \CII,  
[O$\,${\sc i}] (63~$\mu$m) or [O$\,${\sc i}] (145~$\mu$m) 
accurately reflect most of the physical properties of the H$_2$ gas and the total gas of the cloud, whereas molecular tracers ($^{12}$CO or $^{13}$CO) do not.
Synthetic \CII\ line emission studies have also been carried out for star-forming galaxies at high redshift ($z \sim 6$) by e.g.\ \cite{Olsen.etal.2017}, who investigate whether the \CII\ line emission can be used as a star formation tracer at these epochs. 
For this purpose they also study the origin of the \CII\ line emission and find giant molecular clouds to be the source of a majority of the \CII\ emission. \\

Our work builds on 3D hydrodynamical simulations of a molecular cloud. 
In the framework of the SILCC simulations,\footnote{SILCC: Simulating the Life Cycle of molecular clouds, project led by S.Walch, \url{https://hera.ph1.uni-koeln.de/~silcc/} } the evolution of the ISM is modelled self-consistently within a piece of a galactic plane (0.5\,kpc\,$\times$\,0.5\,kpc\,$\times$\,$\pm 5$\,kpc) with initial conditions inspired by the values of the solar neighbourhood \citep{Walch.etal.2015,Girichidis.etal.2016,Gatto.etal.2015,Gatto.etal.2016,Peters.etal.2017}. One important feature of these simulations is their use of simplified chemical network, which calculates the non-equilibrium evolution of hydrogen and carbon species within the gas. Based on the SILCC simulations, \cite{Seifried.etal.2017} carried out zoom-in simulations of the evolution of two molecular clouds and their chemistry using a greatly improved resolution (d$x \geq 0.122$~pc) as part of the SILCC-Zoom project. In these zoom-in simulations, although high resolution is applied within the molecular cloud region, the surrounding galactic environment is retained, albeit at lower resolution. This allows for the galactic environment to influence the growth of molecular clouds via large-scale turbulent flows in a self-consistent manner. \\

In this paper, we post-process the results of one of these zoom-in simulations using the \RADMC\ radiative transfer code \citep{Dullemond.2012.II} in order to produce synthetic \CII\ emission maps.  Our aim is to understand how much \CII\ emission is produced during the formation of a typical molecular cloud and from which regions in the cloud this emission comes. In addition, we briefly discuss the numerical resolution required in order to produce numerical converged predictions for the \CII\ emission. We note that, since we do not model the formation of stars within the growing molecular clouds in this work, we are not able to account for the \CII\ emission produced in PDRs around newly-formed massive stars. Instead, we concentrate on an earlier stage in the lifetime of the cloud, before massive star formation has fully got underway \citep[as observed in e.g. ][]{Beuther.etal.2014}. We focus on analysing the origin of the velocity-integrated \CII\ line emission, deferring any discussion of \CII\ as a kinematic tracer to future work.\\
 
The structure of the paper is as follows. In Section~\ref{sec.Methods}, we briefly describe the method for producing the zoom-in simulations and the radiative transfer post-processing. In Section~\ref{sec.Resolution}, we investigate the impact of the numerical resolution of the simulations on the synthetic \CII\ line emission maps. In Section~\ref{sec.Analysis} we study the origin of the \CII-emitting gas and try to disentangle its 3D structure. Here, to avoid optical depth effects, we treat the \CII\ emission as being optically thin. 
We then analyse the correlation between the total gas, H and C$^+$ column densities and the \CII\ line emission in Section~\ref{sec.XCII}, and close with a discussion and summary of our results in Sections~\ref{sec.Discussion} and \ref{sec.Conclusion}. 

\section{Methods}
\label{sec.Methods}

\subsection{SILCC Zoom-in simulations}\label{sec.zoom}

The simulations of molecular clouds (zoom-ins, SILCC-Zoom project) are based on the SILCC simulations \citep{Walch.etal.2015, Girichidis.etal.2016}. 
The zoom-ins and the applied technique were presented in \citet[][hereafter S17]{Seifried.etal.2017}. 
In the following we briefly describe the numerical methods used. For more details we refer the reader to the aforementioned papers. \\

The simulations are performed with the adaptive mesh refinement code \FLASH\ 4.3 \citep{Fryxell.2000, Dubey.etal.2008} using a magneto-hydrodynamics solver which guarantees positive entropy and density \citep{Bouchut.etal.2007,Waagan09}. The chemical evolution of the ISM is modelled using a simplified chemical network for H$^+$, H, H$_2$, C$^+$, CO, e$^-$, and O \citep{Nelson.Langer.1997,Glover.MacLow.2007.II,Glover.Clark.2012}. We do not include high-temperature shock chemistry, as our simulations do not have sufficient resolution to resolve the hot and thin post-shock cooling zones behind highly supersonic shocks. We assume solar metallicity with elemental abundances of carbon and  oxygen relative to hydrogen given by $1.4 \times 10^{-4}$ and $3.16 \times 10^{-4}$, respectively \citep{Sembach.etal.2000}. The chemical network also follows the thermal evolution of the gas including the most relevant heating and cooling processes. The ISM is embedded in an interstellar radiation field (ISRF) of $G_0 = 1.7$ in units of \cite{Habing.1968}, that is in line with \cite{Draine.1978}. The ISRF is shielded in dense regions according to the surrounding column densities of the total gas, dust, H$_2$, and CO, calculated via the \textsc{TreeRay OpticalDepth} module \citep{Wunsch.etal.2018} based on the \textsc{TreeCol} algorithm \citep{Clark.etal.2012}. This enables us to calculate the photochemical reaction rates and radiative heating rates as well as the dust temperature. We solve the Poisson equation for self-gravity with a tree-based method \citep{Wunsch.etal.2018} and include a background potential due to the pre-existing stellar component in the galactic disc, which is modeled as an isothermal sheet with \mbox{$\Sigma_\mathrm{star}$ = 30 M$_{\sun}$ pc$^{-2}$} and a scale height of \mbox{100 pc} \citep{Spitzer.1942}. \\

The SILCC simulation setup represents a small section of a galactic disc with solar neighbourhood properties and a size of 500~pc $\times$~500~pc~$\times$ $\pm$5~kpc. We apply periodic boundary conditions along the $x$- and $y$-directions and outflow conditions along the $z$-direction, which are Neumann boundary conditions, with the additional requirement that the velocity vector is pointing outwards. Otherwise the velocity is set to zero, thus only allowing for outflow. The initial vertical distribution of the gas has a Gaussian profile with a scale height of 30~pc and a midplane density of 9~$\times$~$10^{-24}$~g~cm$^{-3}$, which  gives us a total gas surface density of \mbox{$\Sigma_\mathrm{gas}$ = 10~M$_{\sun}$~pc$^{-2}$}. The gas around the disc midplane has an initial temperature of 4500~K. Hydrogen in this gas is assumed to start in atomic form, while the carbon is present initially as C$^{+}$. \\

From the start we inject SNe up to \mbox{$t_0 = 11.9$~Myr} with a constant rate of 15~SNe~Myr$^{-1}$ in order to allow a self-consistent three-phase ISM to develop. This SN rate is chosen in agreement with the Kennicutt-Schmidt star formation rate surface density for \mbox{$\Sigma_\mathrm{gas} = 10$~M$_{\sun}$~pc$^{-2}$} \citep{Schmidt.1959,Kennicutt.1998}. Half of the SNe explode within the density peaks, and the other half are randomly placed \citep[mixed driving; see][for more details]{Walch.etal.2015}. For a single SN explosion 10$^{51}$~erg is injected \citep[see][for more details]{Gatto.etal.2015}. \\

The grid resolution up to $t_0$ is 3.9~pc. At $t_0$ we stop the injection of further SNe since we want to study a molecular cloud which is unaffected by SNe in its further evolution. We choose two regions in which molecular clouds (MCs) --- henceforth denoted as ``zoom-ins'', or MC1 and MC2 --- are about to form (see S17 for more details). At this point, the zoom-in regions have typical densities below a few times 10~cm$^{-3}$. We then continue the simulation for another 1.5~Myr and progressively increase the spatial resolution d$x$ in these two regions from the refinement level ``L5'' with d$x = 3.9$~pc to the refinement level ``L10'' with d$x = 0.122$~pc (see table~2 in S17). In the surroundings of the zoom-ins we retain the SILCC simulation box at the lower resolution of 3.9~pc, thus keeping the effect of accretion onto the cloud from the larger-scale, galactic environment. 
After reaching the highest resolution level (L10 with d$x = 0.122$~pc), we continue the simulation at that level. 
As shown in S17, this resolution is sufficient (and required) to accurately model both the chemical and dynamical evolution of the clouds. Furthermore, this resolution allows us to resolve the filamentary substructure of the clouds. \\

Since both MC1 and MC2 have qualitatively similar properties, in particular with respect to their \CII\ emission, we chose to present our analysis for MC2. In particular, we study MC2 at an evolutionary time of $t_{\textrm{tot}} = t_0 + 2\,\textrm{Myr} = 13.9$~Myr.  In S17, we present the CO-to-H$_2$ conversion factor for MC1 and MC2 at $t_{\textrm{tot}} = 13.9$~Myr (see their figs.~17 and 18). The zoom-in region of MC2 has a size of 104 pc $\times$~88~pc~$\times$~71~pc, centred around the midplane. In total, this region contains a gas mass of $8 \times 10^{4}$~M$_{\odot}$. The mass-weighted mean velocity dispersion of MC2, defined for number density thresholds between 30~cm$^{-3}$ and 300~cm$^{-3}$, is about 4~\kms\ (see fig.~10 in S17). \\ 

In order to study the effect of the resolution of the 3D simulation on the obtained \CII\ emission, we consider runs with different maximum resolution levels ranging from ``L5'' to ``L10''. For this we simply stop the progressive refinement approach at a lower maximum resolution level. In this way we guarantee that except for the different maximum resolution, the individual simulations are as close to each other as possible. In Table~\ref{tab.ZoomIn} we list all simulations investigated in this work. In column 1, we list the run names as used by S17, where ``tau'' denotes the further simulation time after starting the refinement procedure, and ``dx'' denotes the minimum cell size. In this work, we call the simulations by their maximum resolution level, i.e. ``L7'' corresponds to run MC2\_tau-0.5\_dx-1.0 (see column 2).\\

\begin{table*}
\begin{minipage}{160mm}
\centering
\caption{List of the zoom-in simulations used here, as named in S17 (first column) and named after their resolution level  (second column). The third column gives the spatial resolution d$x$. Further, we list for the $x-z$ projection the integrated peak intensities and total luminosities \Ltot\, for the opacity affected and the optically thin \CII\ synthetic emission maps (fourth to seventh column, respectively; see Section ~\ref{sec.Resolution}). } 
\label{tab.ZoomIn}
\begin{tabular}{@{}ccccccc@{}}
\hline
Run name in S17 & resolution level & d$x$ & \multicolumn{2}{l}{\CII, opac. aff.} & \multicolumn{2}{l}{\CII, opt. thin} \\
 & & & $I_{\textrm{peak}}$ & \Ltot\ & $I_{\textrm{peak}}$ & \Ltot\ \\
 & & [pc] & [K \kms] &  [L$_{\odot}$]  & [K \kms] &  [L$_{\odot}$]\\
\hline
\hline
MC2\_dx-3.9 & L5 & 3.9 & 1.36 & 353 & 0.01 & 3.00\\
MC2\_tau-0.0\_dx-2.0 & L6 & 2 & 1.90 & 390 & 0.02 & 3.45\\
MC2\_tau-0.5\_dx-1.0 & L7 & 1 & 3.60 & 459 & 0.04 & 4.23\\
MC2\_tau-1.0\_dx-0.5 & L8 & 0.5 & 5.82 & 555 & 0.06 & 5.30\\
MC2\_tau-1.25\_dx-0.24 & L9 & 0.25 & 10.30 & 628 & 0.12 & 6.14\\
MC2\_tau-1.5\_dx-0.12 & L10 & 0.122 & 10.56 & 652 & 0.21 & 6.52\\
\hline
\end{tabular}
\end{minipage}
\end{table*}

\subsection{Radiative transfer simulations}
We post-process the simulations with the radiative transfer code \RADMC\footnote{http://www.ita.uni-heidelberg.de/$\sim$dullemond/software/radmc-3d/} \citep[][version 0.40]{Dullemond.2012.II} to study the signal of the \CII\ emission line. Here we briefly sketch the mode of operation of \RADMC\ and refer to Appendix~\ref{app.RADMC} for a more detailed description, and to the manual of \RADMC\ for even more information. \\

We carry out the radiative transfer for the \CII\ line emission at $\nu_{[\textrm{CII}]} = 1900.537$~GHz ($\lambda_{[\textrm{CII}]} = 157.741$~$\mu$m). Since the \CII\ line emission can become optically thick, we calculate in this paper the \textit{(i)} \CII\ line emission that is influenced by opacity and corresponds to the case of ${[^{12}\textrm{CII}]}$ emission (``opacity affected'' case), and \textit{(ii)} the \CII\ line emission that we treat as if it would be optically thin (``optically thin'' case). For the latter, we modify the number density of $n_{\textrm{C}^+}$ by dividing it by a factor chosen such that the optically thin \CII\ line emission corresponds to the optically thin \CIIdrei\ line emission. Following \cite{Wakelam.Herbst.2008}, the standard isotopic ratio of $^{12}$C$^+$/$^{13}$C$^+ = 67$. The fine structure transition in $^{13}$C$^+$ splits into three \CIIdrei\ hyperfine structure lines $F - F' = 2 - 1, 1 - 0, 1 - 1$, where $F$ refers to the hyperfine structure level, with intensity ratios of 0.625:0.250:0125  \citep[][]{Ossenkopf.etal.2013}. Taking the fraction of the brightest line among these three, we define $n_{\textrm{C}^+, \textrm{opt.~thin}}$ as 
\begin{equation}
\label{eq.noptthin}
n_{\textrm{C}^+, \textrm{opt.~thin}} = n_{\textrm{C}^+} \times \frac{0.625}{67}.
\end{equation}
We take the same frequency to calculate the \CII\ line emission for both the opacity affected and the optically thin case. \\

\RADMC\ solves the radiative transfer equation under the assumption that the profile function is equal for the emission and absorption (complete redistribution). Absorption and emission are calculated internally, using a Gaussian profile function of the line as well as the fractional level populations of the C$^+$ ion. We assume that the line is broadened thermally ($a_{\textrm{therm}}$) and via microturbulence ($a_{\textrm{turb}}$). We set $a_{\textrm{turb}} =  a_{\textrm{therm}}$, with $a_{\textrm{therm}} = \sqrt{\frac{2 k_{\textrm{B}} T}{\mu m_{_{\rm H}}}}$. Here, $T$ denotes the kinetic temperature, $k_{\textrm{B}}$ is the Boltzmann constant, $m_{_{\rm H}}$ is the mass of a hydrogen atom, and $\mu$ is the molecular weight of the emitting $^{12}$C$^+$ particle ($\mu = 12$). We have further investigated whether convergence of our synthetic \CII\ emission maps with increasing resolution is improved if we adopt a value of $a_{\textrm{turb}}$ that depends on the size of the grid cell, following eq.~1 in \citet{Larson.1981}. However, we find that in practice this does not make a significant difference (Appendix~\ref{app.Micro}). \\

In a low density medium the population of the energy levels of C$^+$ does not correspond to the thermal distribution, and thus we do not assume local thermal equilibrium (LTE) in our calculations. In order to calculate the level populations of C$^+$ for the upper and lower levels, we therefore consider collisions with molecular hydrogen (ortho-H$_2$ and para-H$_2$), atomic hydrogen, and electrons \citep{Goldsmith.etal.2012, Lique.etal.2013}. We use the large velocity gradient (LVG) approximation \citep[see][]{Shetty.etal.2011.I, Shetty.etal.2011.II}, i.e. lines\_mode = 3 of \RADMC. We tested several maximum values for the Sobolev length (2~pc and 70~pc) and found that this has a negligible impact on our results (Appendix~\ref{app.EscProb}). In regions of dense gas the level populations approach their LTE values. \\

\begin{figure}
\centering
\includegraphics[width=80mm]{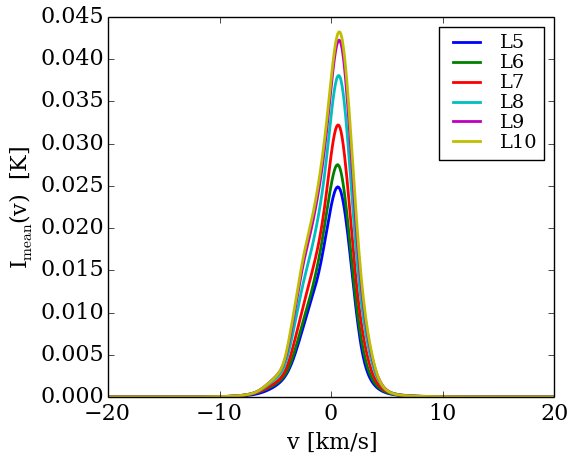}
\caption{Averaged \CII\ spectra for the opacity affected case of the whole synthetic emission map. 201 velocity channels are chosen in a velocity range of $\pm 20$~\kms\ (d$v = 0.2$~\kms) around the rest frequency of $1900.537$\,GHz. Different spatial resolutions are colour-coded (see Table~\ref{tab.ZoomIn}).}
\label{fig.vel}
\end{figure}

In order to capture the contribution of Doppler-shifted emission, we consider a velocity range of $\pm 20$~\kms, centred around $\nu_{\textrm{[CII]}}$. We divide this range into 201 equally spaced velocity channels, corresponding to a spectral resolution of d$v = 0.2$~\kms. The choice of this spectral resolution is inspired by the \textit{Herschel} observations of the \CII\ line emission in Orion \citep[][with d$v = 0.2$~\kms]{Goicoechea.etal.2015}. In Appendix~\ref{app.SR} we test different spectral resolutions.The radiative transfer calculation is done for every pixel in the projected image. The spatial resolution of the synthetic emission maps corresponds to the maximum resolution d$x$, as listed in Table~\ref{tab.ZoomIn}. Figure~\ref{fig.vel} depicts the averaged spectrum of MC2 at different resolution levels (colour-coded). The average spectrum shows the mean intensity calculated over the whole zoom-in region in each velocity channel. Most of the emission comes from velocities between $\pm 10$~\kms. With an increasing resolution level the maximum of the averaged spectrum grows and converges for the higher resolution levels L9 and L10. As described in Section~\ref{sec.Resolution}, the density within the simulation increases with increasing resolution, resulting in higher intensities.  \\

We provide \RADMC\ with the collisional rates and the number densities of C$^+$ and its collisional partners as well as with the gas temperatures in order to carry out the radiative transfer calculation. 
In the following we discuss these input parameters in more detail. \\

\subsubsection{Collisional rates}
\label{sec.CollRates}

The de-excitation rate coefficients of ortho- and para-molecular hydrogen, atomic hydrogen and electrons with C$^+$ are taken from the Leiden Atomic and Molecular Database\footnote{http://home.strw.leidenuniv.nl/$\sim$moldata/} \citep[LAMDA,][]{Schoeier.etal.2005}, and extrapolated to higher temperatures following \cite{Goldsmith.etal.2012}. \RADMC\ converts the de-excitation rates internally into excitation rates as described in Appendix~\ref{app.RADMC}. Since the temperature ranges present in the SILCC simulations are larger than the temperature ranges for which the coefficients are given, we extrapolate the de-excitation rate coefficients to higher temperatures. A detailed description of how we obtain the collisional rates is presented in Appendix~\ref{app.RADMC}. \\

\subsubsection{Number densities}
To obtain information about the number densities within the zoom-in region, we convert the \FLASH\ data to \RADMC\ input data. The information of the \FLASH\ grid structure within the zoom-in region is passed to the \RADMC\ input data by creating an oct-tree structured binary file. This is identical to the original \FLASH\ grid structure. Compared to an uniform grid structure at the highest resolution, the oct-tree structure allows us to reduce the memory requirements. The \RADMC\ input data only include the information of the zoom-in region of $(104 \times 88 \times 71)\,\textrm{pc}^3$. By this we ensure that we study only the emission from the molecular cloud itself and that we are not contaminated by fore- and background emission from unresolved regions. \\

\begin{figure*}
\centering
\includegraphics[width=\textwidth]{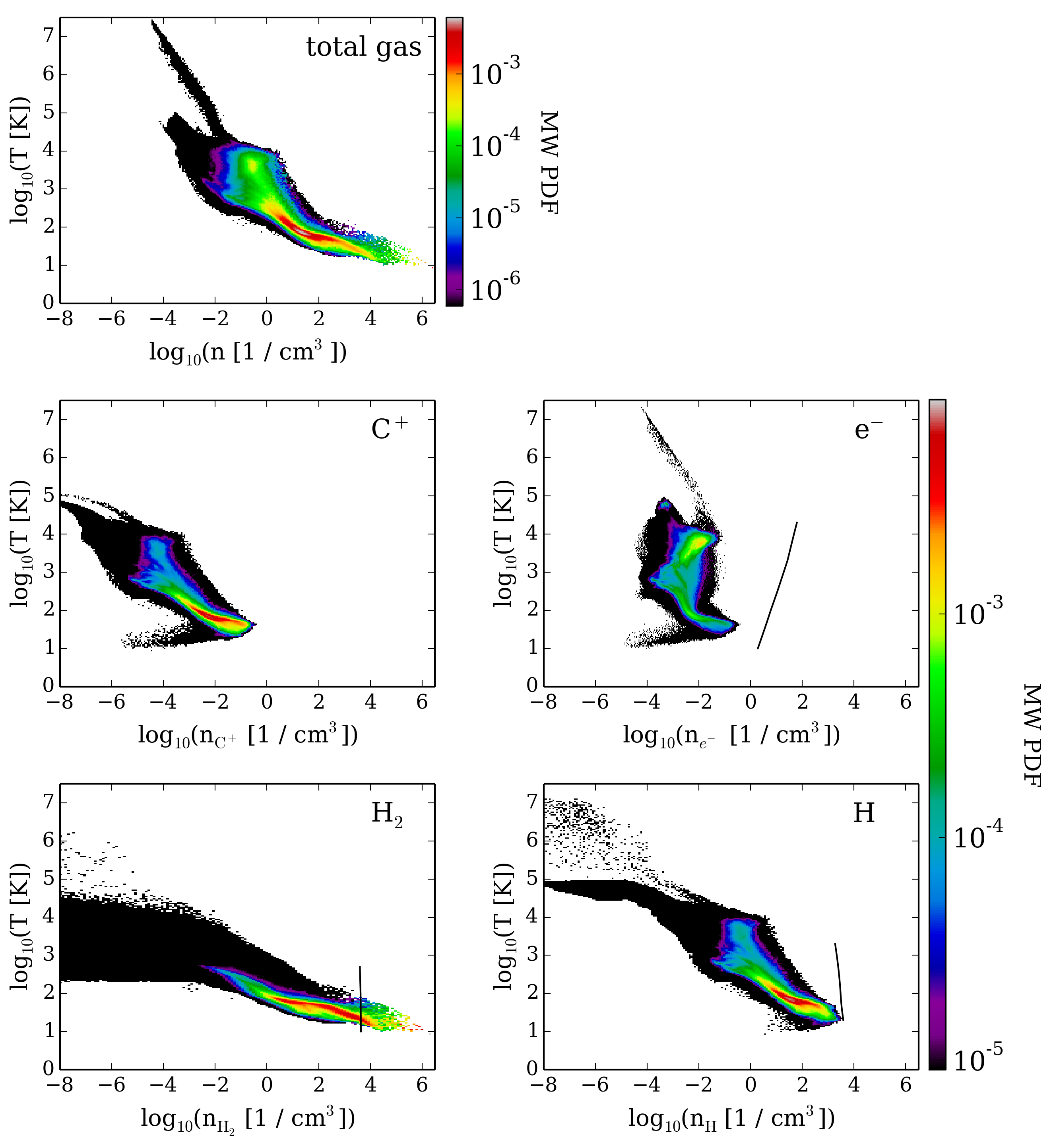}
\caption{Mass-weighted 2D-PDFs as a function of the total gas temperature $T$ and number density of each chemical species in the L10 simulation for the total gas (upper left panel), and for C$^+$ and its primary collisional partners, $e^-$, H$_2$, and H (as indicated in the plots). The colour-coding shows the mass fraction normalized to the total gas mass of the respective species. The black lines indicate the critical densities of each collisional partner. For all but H$_2$ the number densities of the individual collisional partners are below the critical densities. We therefore expect the excitation of C$^+$ to be subthermal in all gas apart from regions which are dominated by H$_2$.}
\label{fig.ZoomIn_Phaseplots_separate}
\end{figure*}

Figure~\ref{fig.ZoomIn_Phaseplots_separate} illustrates the mass-weighted 2D probability distribution function (2D-PDF, equivalent to a two-dimensional histogram)\footnote{In a two-dimensional mass-weighted PDF we show a parameter space of two quantities, and plot how the mass is distributed within this parameter space. The colour-coding indicates the fraction of mass.} of the L10 simulation for the total gas (upper left panel), as well as for C$^+$, $e^-$, H$_2$ and H (lower four panels). The 2D-PDFs show the distributions of the gas as a function of the number density of each chemical species and the gas temperature, with bin sizes of $\Delta \textrm{log}_{10}(T) = 0.03$, $\Delta \textrm{log}_{10}(n\;[\textrm{cm}^{-3}]) = 0.05$, $\Delta \textrm{log}_{10}(n_{\textrm{C}^+}\;[\textrm{cm}^{-3}]) = 0.06$, $\Delta \textrm{log}_{10}(n_e\;[\textrm{cm}^{-3}]) = 0.02$, $\Delta \textrm{log}_{10}(n_{\textrm{H}_2}) = 0.1$ and $\Delta \textrm{log}_{10}(n_{\textrm{H}}\;[\textrm{cm}^{-3}]) = 0.07$. 
We derive the number densities of the chemical species in the following way: \\

\noindent{\sc $n_{\textrm{C}^+}$:} 
The simplified chemical model used in the SILCC simulations does not include higher ionization states of carbon such as C$^{2+}$ or C$^{3+}$. If we simply were to take the C$^+$ abundance from the simulation data, we would therefore overestimate the C$^+$ number density in gas warmer than $T \sim$20\,000~K. To correct for this, we post-process the C$^+$ number densities obtained from the SILCC simulations by applying an ionization correction factor (ICF). The ICF is taken from \citet[][fig.~3 therein]{Sutherland.etal.1993} and assumes collisional ionization equilibrium (CIE). By multiplying the C$^+$ number density from the simulation with the ICF, it is reduced in gas with $T > 20$\,$000$~K and for $T > 200$\,$000$~K all carbon is in a multiply ionized state. The resulting C$^+$ number density is denoted with $n_{\textrm{C}^+}$. 
We expect this procedure to have a minor impact on the resulting synthetic emission maps, since the density of the hot gas is very low. The resulting 2D-PDF for C$^+$ is shown in the middle left panel of Fig.~\ref{fig.ZoomIn_Phaseplots_separate}. \\

\noindent{\sc $n_{e^-}$:} 
The middle right panel of Fig.~\ref{fig.ZoomIn_Phaseplots_separate} shows the 2D-PDF for the electrons. We assume that the gas is electrically neutral overall, which means that the number density of free electrons must balance the total number density of ions. Therefore, the two ions we track in our chemical model, H$^{+}$ and C$^{+}$, determine the electron number density. Since the fraction of carbon is very low compared to the fraction of hydrogen in the simulations, H$^{+}$ is the main donor of electrons in highly ionized regions. In cold gas, the H$^{+}$ abundance can become very small; in these conditions, C$^{+}$ provides most of the free electrons. We therefore have 
\begin{equation}
n_e = n_{\textrm{H}^+} + n_{\textrm{C}^+}.
\end{equation} 
This expression neglects the contribution to $n_{e}$ from other metals with ionization potentials below 13.6~eV (e.g.\ Si, Mg, Fe), but in regions where C$^{+}$ is abundant, this is at most a small correction. In dense gas with a low C$^{+}$ abundance, these low ionization potential metals can contribute a much larger fraction of the total electron abundance, but these regions produce little \CII\ emission.

\noindent{\sc $n_{\textrm{H}_2}$:} 
The number density of H$_2$ is taken directly from the simulation (Fig.~\ref{fig.ZoomIn_Phaseplots_separate}, lower left panel). Additionally, we distinguish between the two nuclear spin states of H$_2$, in which the spins of the nuclei are parallel ($I = 1$, \mbox{ortho-H$_2$}) or antiparallel ($I = 0$, \mbox{para-H$_2$}). When many different rotational levels are populated (e.g.\ at high temperatures and densities), the equilibrium ortho-to-para ratio is given by the ratio of the nuclear statistical weights \mbox{$g_{I} = 2I + 1$} and hence, it is 3:1. At low temperatures ($T \lesssim 150$~K), where only the $J=0$ and $J=1$ rotational levels are populated, the ortho-to-para ratio is given instead by \citep{Rachford.etal.2009} 
\begin{equation}
\frac{n( \textrm{ortho-H} _2)}{n( \textrm{para-H} _2)} = 9 \times e^{-171 \textrm{\,K}/T_{\textrm{rot}}},
\end{equation}
where $T_{\textrm{rot}}$ is the rotational temperature. We here use the kinetic temperature $T_{\textrm{kin}}$ as an approximation for $T_{\textrm{rot}}$. Although this is a rough estimate, it is sufficient for our purpose, since the collisional rates for \mbox{ortho-H$_2$} and \mbox{para-H$_2$} are similar and distinguishing between them has only a minor impact on the \CII\ line emission. \\

\noindent{\sc $n_{\textrm{H}}$:} 
The number density of atomic hydrogen is taken directly from the SILCC simulation. \\

The black lines in the 2D-PDFs of the collisional partners in Fig.~\ref{fig.ZoomIn_Phaseplots_separate} indicate the critical densities. The critical density (\cite{Tielens.Hollenbach.1985} 
or \cite{Draine.2011}, chapter 17.2) is defined as the ratio of the spontaneous emission rate --- described by the Einstein coefficient $A_{ul}$ --- and the collisional de-excitation rate coefficient $R_{\textrm{ul}}$ --- with 
\begin{equation}
n_{{\rm crit}} = \frac{A_{ul}}{R_{\textrm{ul}}}.
\end{equation}
Since the size of $R_{\rm ul}$ depends on the collision partner, we obtain a different critical density for each main collision partner. If the number density of any of these collision partners exceeds the corresponding critical density, then collisional de-excitation occurs more rapidly on average than radiative de-excitation. In these conditions, the C$^{+}$ level populations approach their LTE values. From Fig.~\ref{fig.ZoomIn_Phaseplots_separate}, we see that in some regions (although not everywhere), H$_{2}$ has a number density greater than $n_{\rm crit, H_{2}}$. On the other hand, hydrogen atoms and electrons always have number densities less than their corresponding critical densities. Therefore, we expect C$^{+}$ to be subthermally excited in regions dominated by atomic or ionized hydrogen, but to be thermally excited in dense molecular regions.

\section{Synthetic emission maps at different resolutions}
\label{sec.Resolution}

\begin{figure*}
\centering
\includegraphics[width=0.8\textwidth]{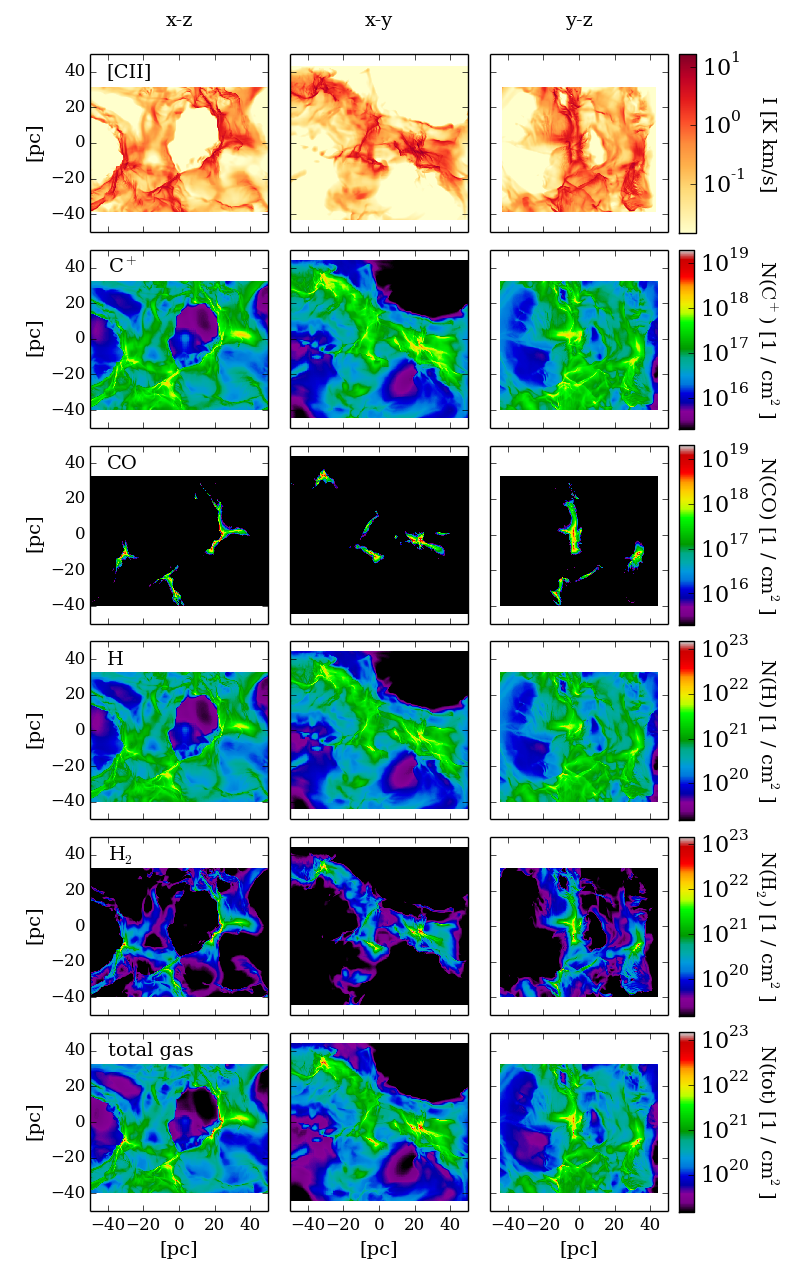}
\caption{Synthetic \CII\ emission maps (opacity affected case, first row) of MC2 in the zoom-in region of $(104 \times 88 \times 71)\,\textrm{pc}^3$ at the resolution level L10 shown for all three projections, as indicated above the images. The following rows present the C$^+$, CO, H, H$_2$, and total gas column density of the cloud, respectively. The \CII\ emission mostly traces the atomic hydrogen column density.}
\label{fig.ZoomIn_Intro}
\end{figure*}

With the method described above we obtain channel maps for each frequency in the velocity range $\pm 20$~\kms\ for a snapshot of MC2 at an evolutionary time of $t_{\textrm{tot}} = t_0 + 2\,\textrm{Myr} = 13.9$~Myr. We integrate the intensity over the whole velocity range and refer to the resulting maps as synthetic emission maps. In the first row of Fig.~\ref{fig.ZoomIn_Intro}, the opacity affected \CII\ synthetic emission maps are shown for MC2 at the maximum refinement level L10. The integrated intensity is colour-coded and presented in units of [K~\kms]. The three columns show the cloud in different projections (as indicated above the columns). The second to sixth rows show the C$^+$, CO, H, H$_2$ and total gas column densities of the cloud.  H$_2$ is present in almost all regions of the molecular cloud, and H in the envelopes of these regions. CO forms at even higher visual extinction and is therefore only present in the dense parts of the molecular cloud \citep[cf.][S17]{Roellig.etal.2007, Glover.etal.2010, Smith.etal.2014, Duarte-Cabral.Dobbs.2016, Xu.etal.2016}. Since no atomic carbon is considered in the chemical network, we might slightly overestimate the production of CO and consequently underestimate the amount of C$^+$ present in dense gas \citep{Glover.Clark.2012}. C$^+$, as the left-over from carbon that is not in the form of CO, is distributed in most parts of the cloud and is only reduced in dense regions, where CO is present. The \CII\ line emission recovers in general the density structure of the cloud. However, in regions with high column densities,  $N \gtrsim  10^{23}$~cm$^{-2}$, CO becomes abundant, C$^+$ is reduced and therefore also the \CII\ line emission drops. One example for this can be seen in the center of the synthetic emission map for the $y-z$ projection (third column). \\

\begin{figure*}
\centering
\includegraphics[width=\textwidth]{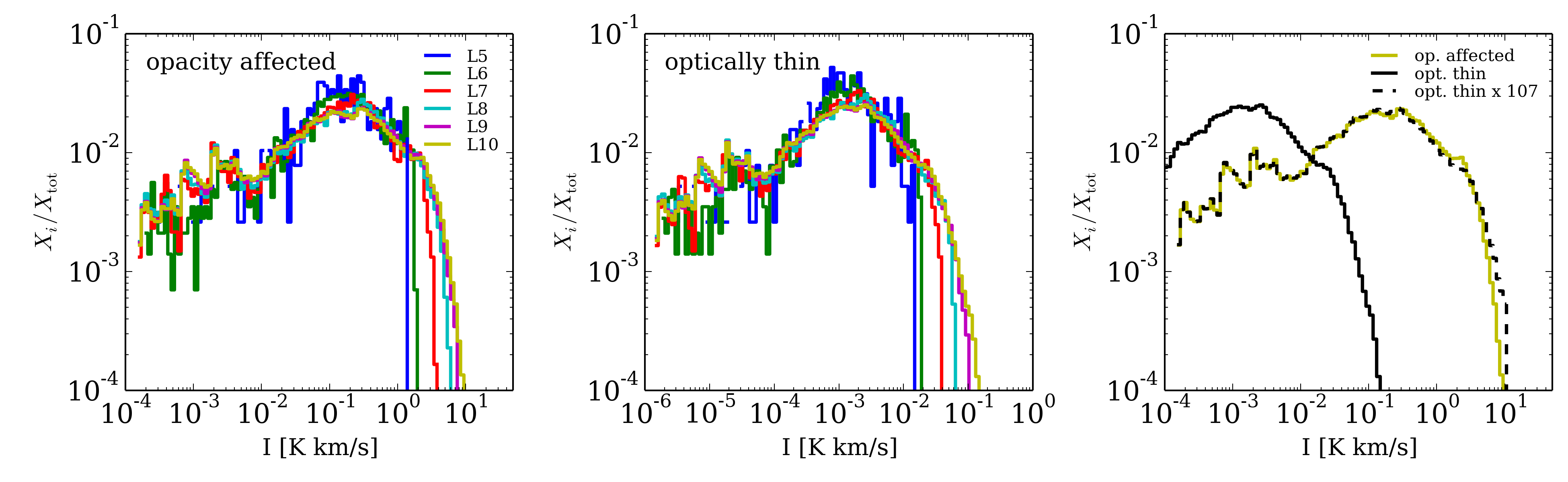}
\caption{Histogram of the integrated \CII\ intensity for the $x-z$ projection for the opacity affected case (left) and the optically thin case (middle). $X_i / X_{\textrm{tot}}$ denotes the fraction of pixels in a given intensity bin. The different resolutions are colour-coded (see Table~\ref{tab.ZoomIn}). The histograms converge for a resolution level higher than L9. In the right panel, we compare the histograms of the opacity affected and optically thin \CII\ line emission. To bring the intensities to a comparable range, we scale the optically thin emission by a factor of 107, according to Eq.~\eqref{eq.noptthin}. The histograms are identical, except in the range for $I_{[\textrm{CII}]} \geq 0.1$~K~\kms, where the emission is optically thick.}
\label{fig.IntPDF}
\end{figure*}

In Table~\ref{tab.ZoomIn} we list the maximum intensities as well as the luminosities of the maps, and in Fig.~\ref{fig.IntPDF} we show the histogram of the integrated \CII\ intensities (with a bin size of $\Delta \textrm{log}_{10}(I\;[\textrm{K\,km\,s}^{-1}]) = 0.05$) for the $x-z$ projection for the opacity affected emission (left) and the optically thin emission (middle). The distributions demonstrate that the maximum integrated intensity increases with refinement level. The distributions of all maps slowly converge for L9 and higher, in particular for the opacity affected case. Otherwise the optically thin case, which is a proxy for the \CIIdrei\ emission, resembles the opacity affected case. These findings are in line with the averaged spectrum in Fig.~\ref{fig.vel}, where the peak intensity increases with the higher resolution. In the right panel of Fig.~\ref{fig.IntPDF} we directly compare the histograms of the opacity affected and optically thin \CII\ integrated intensities. Note, that in order to compare the intensities, we scaled the optically thin \CII\ line emission by a factor of 107, in line with Eq.~\eqref{eq.noptthin}. Only for intensities of $I_{[\textrm{CII}]} \geq 0.1$~K~\kms, where the emission becomes partially optically thick, some minor differences are visible. Further convergence studies for the synthetic \CII\ emission maps are presented in Appendix~\ref{app.ConvMaps}. There we show the synthetic emission maps for the opacity affected and optically thin \CII\ line emission in Figs.~\ref{fig.Mom0_Res} and  \ref{fig.Mom0_Res_CII13}, respectively. In addition we analyse the convergence between the resolution levels and find that the synthetic emission maps of L9 and L10 converge to within 0.5\%, as calculated over the whole map. \\

\begin{figure*}
\centering
\includegraphics[width=\textwidth]{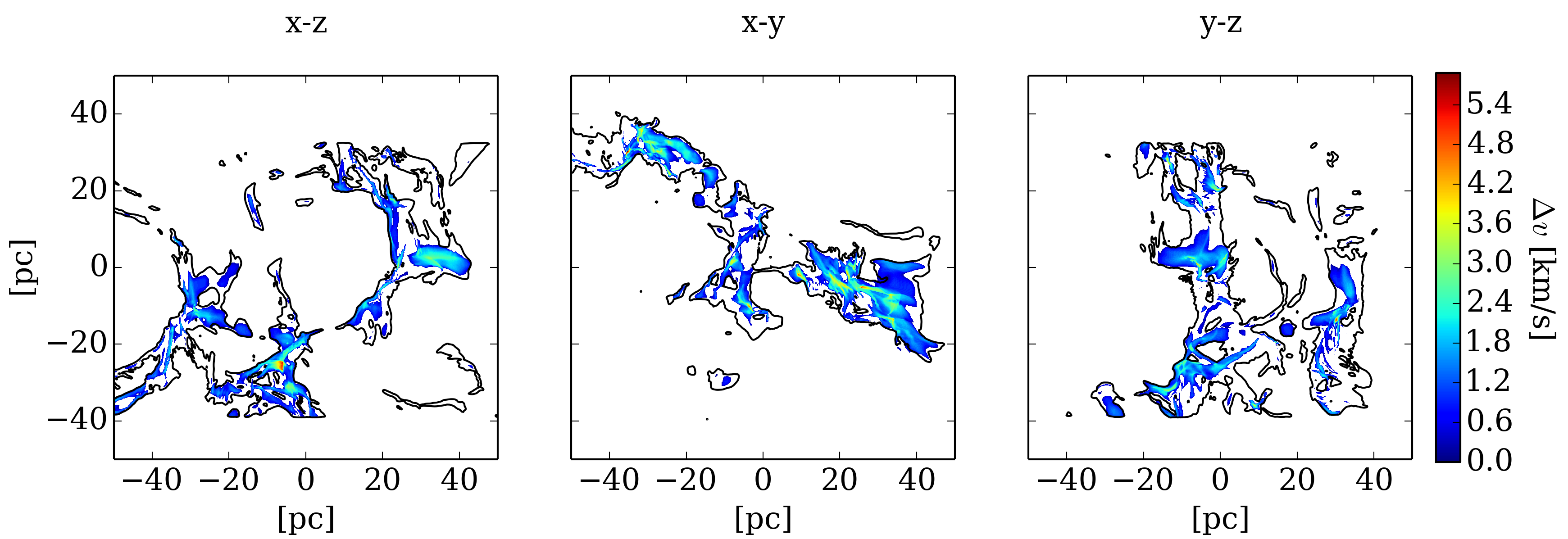}
\caption{Maps of the different projections for L10, showing where the \CII\ line emission becomes optically thick ($\tau_{[\textrm{CII}]} \geq 1$) in at least one velocity channel. The colour-coding shows the width in velocity space over which the emission is optically thick. The black contours mark the observable area for which $I_{[\textrm{CII}]} \geq 0.5$~K~\kms. About 39\%, 34\% and 47\% of the observable area is optically thick for the projections $x-z$, $x-y$ and $y-z$, respectively.}
\label{fig.Tau}
\end{figure*}

Further, we investigate the optical depth for the \CII\ line emission. 
With \RADMC\ we calculate whether the emission along a line of sight becomes optically thick ($\tau_{[\textrm{CII}]} \geq 1$) for a specific wavelength. We find optically depths up to $\tau \leq 10$ in our simulation. The sum over all the channels that contain optically thick emission along one line-of-sight defines the velocity range $\Delta v$, which is colour-coded in Fig.~\ref{fig.Tau}. In all projections around 10\% of the total area is optically thick, accounting for 45\%, 54\% and 40\% of the total luminosity in the $x-z$, $x-y$ and $y-z$ projections, respectively. However, large portions of the synthetic \CII\ emission maps show low integrated intensities, as can be seen in Fig.~\ref{fig.ZoomIn_Intro} (first row). We assume that pixels with integrated intensities $I_{[\textrm{CII}]} \geq 0.5$~K~\kms\ are observable.\footnote{This number is roughly comparable with the \mbox{$3 \sigma$} detection limits of \cite{Velusamy.Langer.2014} and \cite{Croxall.etal.2012}. For example, in the work of \cite{Croxall.etal.2012}, \CII\ maps from nearby spiral galaxies are studied, observed with the PACS instrument on board of the \textit{Herschel} satellite. They report a \mbox{$3 \sigma$} detection limit of approximately $0.04 \times 10^{-7}$~W~m$^{-2}$~sr$^{-1} = 4 \times 10^{-6}$~erg~s$^{-1}$~sr$^{-1}$~cm$^{-2}$. In temperature units, this is equivalent to the integrated intensity 0.57~K~\kms. Better sensitivities can be achieved for observations of \CII\ concentrating on one or a few pointings, but a value of $0.5$~K~\kms\  remains representative of the limiting sensitivity that can be achieved in extended maps of \CII\ with current facilities.} In the following, we call the area with integrated intensities $I_{[\textrm{CII}]} \geq 0.5$~K~\kms\ ``observable area'', as opposed to the ``total area'' of the map. By accounting now only for pixels fulfilling $I_{[\textrm{CII}]} \geq 0.5$~K~\kms\ (black contours), about 21\%, 16\% and 27\% of the total map are observable in the $x-z$, $x-y$ and $y-z$ projection. This observable area contains about 80\% of the total luminosity. If we consider only this observable region, we find that about 39\%, 34\% and 47\% of the observable area is optically thick in the three projections. \\

\section{Origin of the \CII\ line emission }
\label{sec.Analysis}

We aim to study how much of the \CII\ line emission originates from gas in different temperature and density regimes. In Fig.~\ref{fig.Tau} we have shown that the \CII\ line emission is affected by optical depth effects in 10\% of the total area and $\sim$40\% of the observable area ($I_{[\textrm{CII}]} \geq 0.5$~K~\kms), respectively. Hence, we analyse the origin of the  \CII\ line emission for both, the opacity affected and the optically thin cases, and for the total as well as the observable area. With the optically thin tracer we eliminate optical depth effects and demonstrate that our results are not biased by them. Essentially, we find that both analyses yield qualitatively similar results. This behaviour is in line with the findings of \cite{Goldsmith.etal.2012}, where the sufficiently weak integrated \CII\ line emission is proportional to the C$^+$ column density along a line of sight. \cite{Goldsmith.etal.2012} call this behaviour ``effectively optically thin''. We present the differences between the opacity affected and optically thin \CII\ line emission on the example of the origin of the emission as a function of the gas temperature (see Sec.~\ref{sec.Analysis_Temp}) and carry out all following analyses with the optically thin \CII\ line. The results for all analyses are summarized in Tables~\ref{tab.CumInt} and \ref{tab.CumInt_CIIzwei}. \\

\begin{figure}
\centering
\includegraphics[width=80mm]{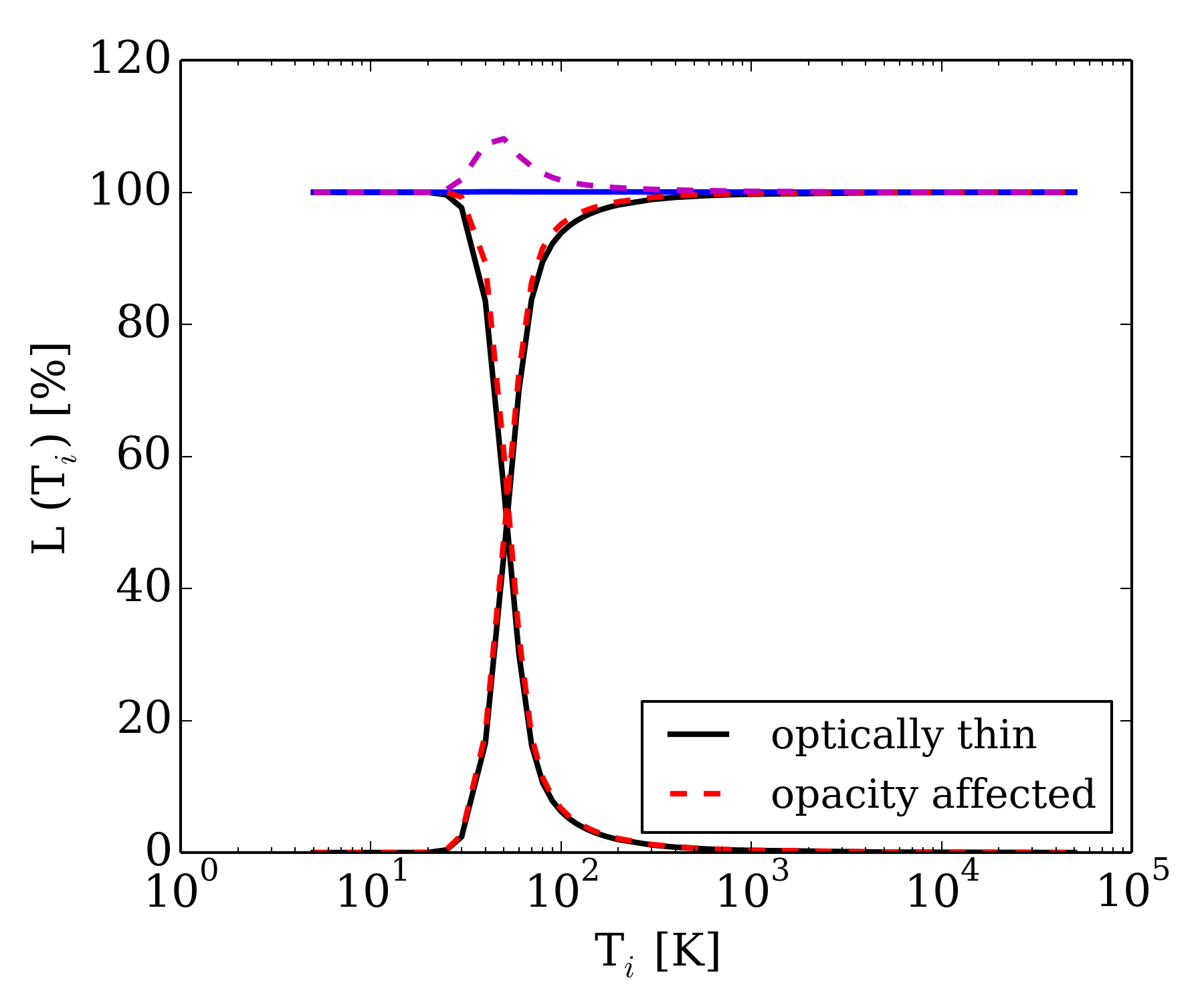}
\caption{Cumulative distributions of the \CII\ luminosity arising from regions with $T \le T_i$ and $T > T_i$ for the opacity affected (dashed red line) and the optically thin (solid black line) case. Summing both lines should give the total luminosity. This is fulfilled for the optically thin \CII\ line emission (solid blue line), but for the optically thick case the total adds up to more than 100\% (dashed magenta line). }
\label{fig.O_12CII_13CII}
\end{figure}

\begin{table}
\centering
\begin{minipage}{80mm}
\caption{Summary of the parameter space of the \CII\ emitting gas, analysed for the optically thin \CII\ line emission. 
We give the values of \Tkin, $n$, the fractional abundance of H$_2$ ($x_{\textrm{H}_2}$), and the visual extinction (A$_{\rm V}$) at which 25\%, 50\% (median) and 75\% of the total luminosity are reached. The corresponding plots are shown in Figs.~\ref{fig.ZoomIn_Origin_T} and \ref{fig.ZoomIn_Origin_07}. The results are presented for the whole map and the observable region (where $I_{[\textrm{CII}]} \geq 0.5$~K~\kms). } 
\label{tab.CumInt}
\begin{tabular}{@{}llccc@{}}
\hline
\CII,   & taken pixels & \multicolumn{3}{c}{$L / L_{\textrm{tot}}$} \\
opt. thin & &  $ \geq 25$\% & $\geq 50$\% & $\geq 75$\% \\
\hline
\hline
\Tkin\ & total & 43~K & 52~K & 64~K \\
 & observable & 41~K & 48~K & 57~K \\
\hline
$n$ & total & 53~cm$^{-3}$ & 166~cm$^{-3}$ & 438~cm$^{-3}$ \\
 & observable & 103~cm$^{-3}$ & 247~cm$^{-3}$ & 524~cm$^{-3}$ \\
 \hline
$x_{\textrm{H}_2}$ & total & 0.08 & 0.14 & 0.22 \\
 & observable & 0.11 & 0.17 & 0.23 \\
 \hline
A$_{\rm V}$ & total & 0.50 & 0.68 & 0.91 \\
 & observable & 0.60 & 0.76 & 0.96 \\
  \hline
\end{tabular}
\end{minipage}
\end{table}

\begin{table}
\centering
\begin{minipage}{80mm}
\caption{Same as Table~\ref{tab.CumInt}, but analysed for the opacity affected \CII\ line emission. The values of the parameters are similar to the optically thin \CII\ line emission. However, note that the results of the opacity affected \CII\ line emission overestimate the total emission. This table is included here for the sake of completeness. }
\label{tab.CumInt_CIIzwei}
\begin{tabular}{@{}llccc@{}}
\hline
\CII,  & taken pixels & \multicolumn{3}{c}{$L / L_{\textrm{tot}}$} \\
opac. aff. & &  $ \geq 25$\% & $\geq 50$\% & $\geq 75$\% \\
\hline
\hline
\Tkin\ & total & 44~K & 53~K & 65~K \\
 & observable & 40~K & 47~K & 55~K \\
\hline
$n$ & total & 62~cm$^{-3}$ & 193~cm$^{-3}$ & 490~cm$^{-3}$ \\
 & observable & 131~cm$^{-3}$ & 292~cm$^{-3}$ & 610~cm$^{-3}$ \\
 \hline
$x_{\textrm{H}_2}$ & total & 0.08 & 0.13 & 0.20 \\
 & observable & 0.11 & 0.16 & 0.22 \\
 \hline
A$_{\rm V}$ & total & 0.48 & 0.64 & 0.84 \\
 & observable & 0.57 & 0.72 & 0.90 \\
  \hline 
\end{tabular}
\end{minipage}
\end{table}

\subsection{Temperature dependence}
\label{sec.Analysis_Temp}

\begin{figure*}
\centering
\includegraphics[width=\textwidth]{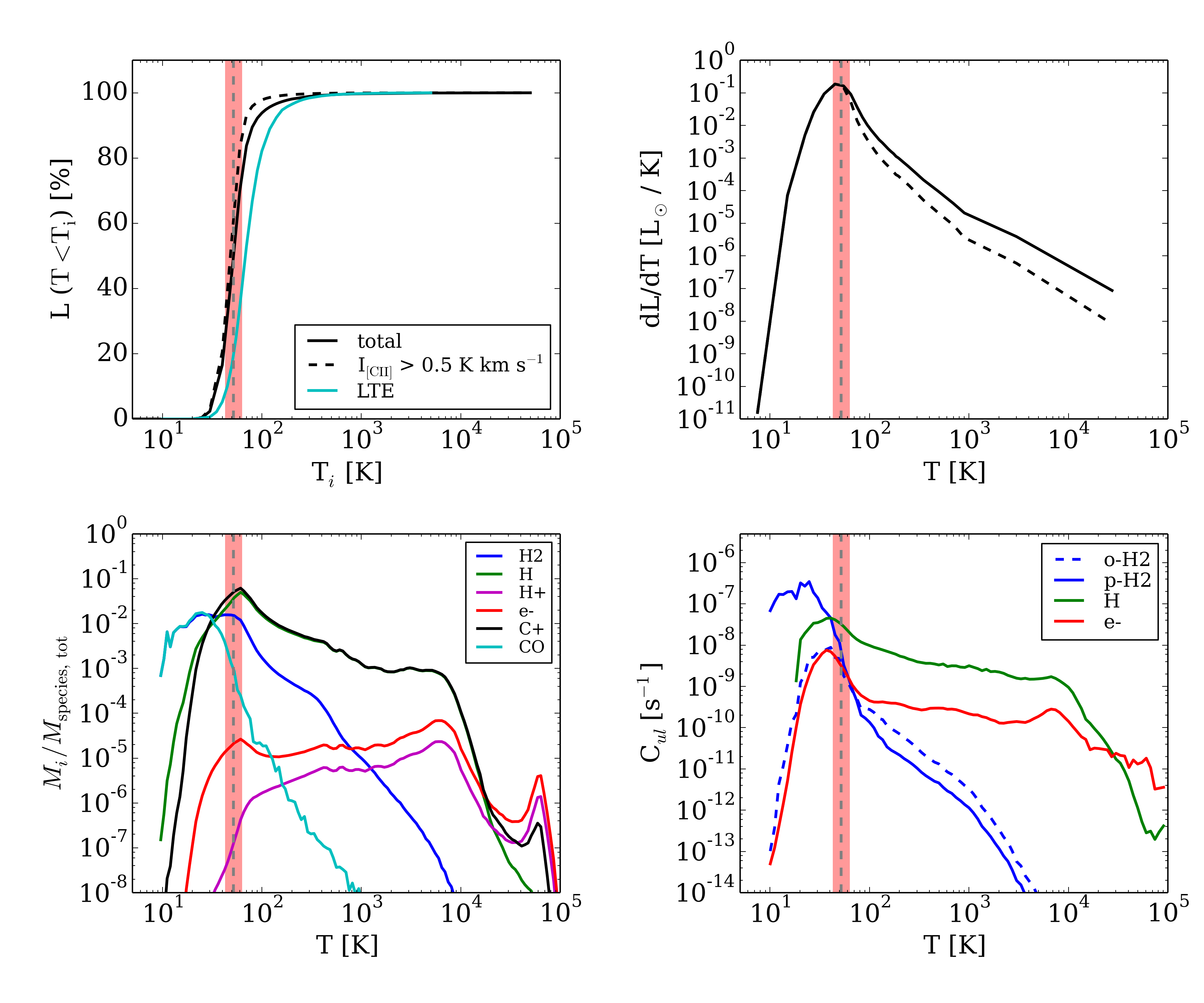}
\caption{\textit{Upper row, left:} Plot of the optically thin \CII\ line luminosity originating from gas with $T \le T_i$ for the total map (solid line) and the assumed observable region (dashed line). The cyan line represents the expectation if all of the C$^+$ ions were in LTE. The red shaded area and the grey, dashed vertical line mark the interquartile range and the median of the distribution derived for the total map. 
\textit{Upper row, right:} The slope of the cumulative luminosity distribution, derived by taking the derivative of the luminosity with respect to the temperature. Most of the emission comes from $40 \lesssim T \lesssim 60$~K.
\textit{Bottom row, left:} Mass-weighted temperature distributions of the included species. $M_i / M_{\textrm{species,~tot}}$ indicates the mass fraction of every species with respect to the total hydrogen, carbon or electron mass, respectively. All collisional partners (H$_2$, H, e$^-$) coexist in the gas from which most of the \CII\ line emission stems, although H is the most abundant one, followed by H$_2$.
\textit{Bottom row, right:} Mass-weighted excitation coefficients for every species (Eq.~\eqref{eq.Cul}) as a function of the temperature. For $T < 40$~K collisions with H$_2$ dominate, whereas for $T > 40$~K collisions with H contribute most to the \CII\ line emission. }
\label{fig.ZoomIn_Origin_T}
\end{figure*}

\begin{figure*}
\centering
\includegraphics[width=\textwidth]{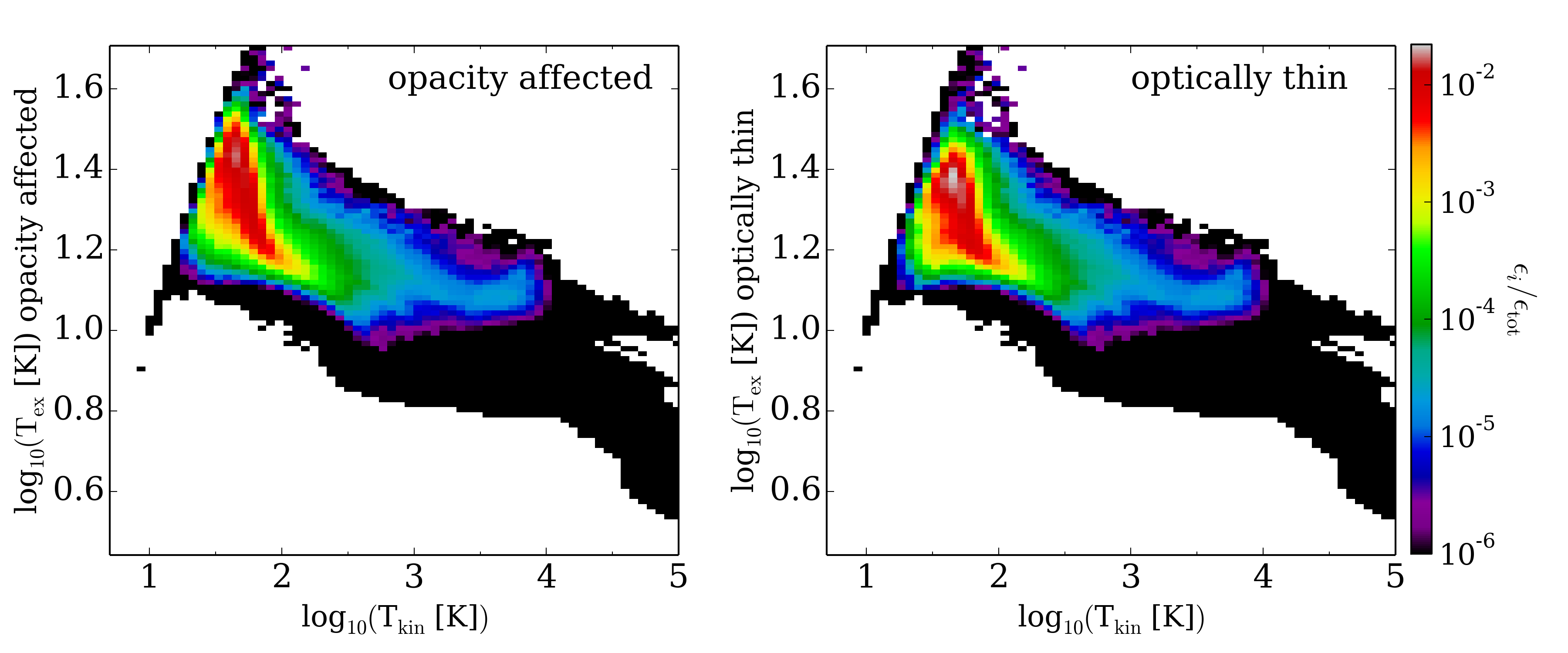}
\caption{Emissivity-weighted 2D-PDFs as a function of the kinetic temperature \Tkin\ and the excitation temperature \Tex\ for the opacity affected \CII\ (left) and the optically thin \CII\ line emission (right). \Tex\ is calculated internally by \RADMC\ and determines the level populations of C$^+$. The colour-coding shows the emissivity from the gas. The identity $T_{\textrm{kin}} = T_{\textrm{ex}}$ can be recognised as the straight upper left boundary of the populated area. }
\label{fig.Tex}
\end{figure*}

From the simulation we know the temperature in every cell and can select cells up to a certain value $T_i$. We calculate the synthetic emission maps for each regime $i$, for which we include gas with $T_{\textrm{kin}} \le T_i$. We increase $T_i$ from \mbox{10~K to $10^5$~K}. In addition, we perform the inverse analysis by selecting all cells with $T_{\textrm{kin}} >T_i$. In the ideal case of an optically thin tracer, the emission for  $T \le T_i$ and $T > T_i$ adds up to the total emission in the map. In Fig.~\ref{fig.O_12CII_13CII} we show how much of the opacity affected \CII\ line emission (red, dashed lines) and the optically thin \CII\ line emission (black, solid lines) originates from gas with $T \le T_i$ (growing from left to right) and $T > T_i$ (dropping off from left to right). If we add up the emission for $T \le T_i$ and $T > T_i$ for the opacity affected \CII\ (dashed, magenta line), we see that the total emission is overestimated by about 10\% for values of $T_i$ between 30~K and 100~K. This occurs because when we include only part of the full range of temperatures, we change the C$^{+}$ optical depth and hence the fraction of the emission able to reach the observer. However, for the optically thin \CII\ emission the contributions from $T \le T_i$ and $T > T_i$ recover 100\% of the total emission (blue, solid line) for all temperature thresholds. Otherwise, the local emission properties are basically the same for the opacity affected and optically thin \CII\ line emission. \\

In Fig.~\ref{fig.ZoomIn_Origin_T} we show the cumulative distribution of the optically thin \CII\ luminosity as function of the temperature of the emitting gas (in the upper left panel), and how much of the emission stems from the different temperature regimes (in the upper right panel). The latter is calculated as the first derivative of the cumulative plot. The solid lines present the analysis for the whole emission map, the dashed lines only for the observable region (where $I_{[\textrm{CII}]} \geq 0.5$~K~\kms). Below 20~K the \CII\ line emission is negligible. For both the analysis with all pixels and only the observable pixels of the synthetic emission maps, almost all of the \CII\ line emission originates from 20~K $\leq T \leq 100$~K. In Table~\ref{tab.CumInt} we list the temperatures at which 25\%, 50\% (median) and 75\% of the total luminosity are reached. In Fig.~\ref{fig.ZoomIn_Origin_T} the red shaded area marks the interquartile range, i.e.\ the difference between the 25$^{\mathrm{th}}$ and the 75$^{\mathrm{th}}$ percentile, for the cumulative plot when taking all pixels into account. The results for the optically thin and opacity affected \CII\ line are similar (see Table~\ref{tab.CumInt_CIIzwei}), with a median ($L/L_{\textrm{tot}} \geq 50$\%) around 50~K. \\

The cyan line shows the LTE prediction for the \CII\ line emission, that is proportional to
\begin{equation}
\label{eq.Ltheo}
L_{\textrm{LTE}} \propto h \nu A_{ul} n_{\rm{C}^+} \frac{g_{u}}{g_{l}} \frac{1}{Z(T)} e^{-\frac{h \nu_{ul}}{k_B T}},
\end{equation}
with the Planck constant, $h$, the Einstein coefficient, $A_{ul}$, the number density of $\textrm{C}^+$, $n_{\rm{C}^+}$, the statistical weights, \mbox{$g_u = 4$} and \mbox{$g_l = 2$}, the Boltzmann constant, $k_B$, the temperature $T$, and the partition function $Z(T)$ defined as 
\begin{equation}
Z(T) = 1 + \frac{g_{u}}{g_{l}} e^{-\frac{h \nu_{ul}}{k_B T}}.
\end{equation} 
LTE would be obtained at high densities. 
The difference between the LTE curve and the curves from the simulations indicates that most of the \CII\ emission is subthermal, in agreement with the fact that most of the gas in the simulation is below the critical density. As discussed already in \cite{Goldsmith.etal.2012}, a large part of the \CII\ emission in the Galaxy stems from thin, subthermally excited gas. Warmer gas is typically less dense than cooler gas, and hence has level populations further from their LTE values. As a result, we obtain a larger fraction of the total \CII\ emission from cooler gas and a smaller fraction from warmer gas than we would expect if the emission were thermal throughout. \\

We compare the available collisional partners as a function of temperature in the lower left panel of Fig.~\ref{fig.ZoomIn_Origin_T}. The temperature bin size is $\Delta\textrm{log}_{10}(T\;\textrm{[K]}) = 0.32$. We show the mass-weighted distributions of the included chemical species. The hydrogen species (H, H$_2$, and H$^+$) are normalized to the total hydrogen mass of $M_{\textrm{H, tot}} = M_{\textrm{H}_2} + M_{\textrm{H}} + M_{\textrm{H}^+}$, and the carbon species (CO and C$^+$) are normalized to the total mass of carbon-bearing species of $M_{\textrm{C, tot}} = M_{\textrm{CO}} + M_{\textrm{C}^+}$. The free electrons are weighted by the mass of electrons that could potentially be free electrons: in our network a carbon species can give one electron, every hydrogen atom can give one, and thus we weight the mass of free electrons by $M_{\textrm{e, tot}} = m_e \times (2n_{\textrm{H}_2} + n_{\textrm{H}} + n_{\textrm{H}^+} + n_{\textrm{CO}} + n_{\textrm{C}^+}) \times d\textrm{V}$. In this way, one can directly read off the relative abundance of the considered collisional partners, i.e. H, H$_2$ and free electrons, for each temperature. Within the interquartile range from which 50\% of the \CII\ line emission originates, atomic hydrogen is the most abundant collisional partner, followed by H$_2$. The amount of free electrons is negligible. Also, over the entire temperature regime the distribution of the C$^+$ ions is aligned with the distribution of atomic hydrogen. Less than 1\% of the C$^+$ mass is at $T \leq 20$~K. About 73\% of the total C$^+$ mass is present in gas at temperatures between 20~K and 100~K. \\

In the bottom right panel of Fig.~\ref{fig.ZoomIn_Origin_T} we study the mass-weighted excitation rates for the chemical species as a function of kinetic temperature. Below 40~K, collisions with \mbox{para-H$_2$} dominate. For temperatures \mbox{$T \geq 40$~K} atomic hydrogen becomes the dominant collisional partner, coinciding with the temperature range from which 50\% of the emission originates, as marked with the red-shaded area. Although H remains the dominant chemical component for higher temperatures \mbox{($T > 100$~K)}, the further contribution to the total \CII\ line emission is low because the density of the gas is small at high temperatures (see upper panels). \\

\begin{figure}
\centering
\includegraphics[width=80mm]{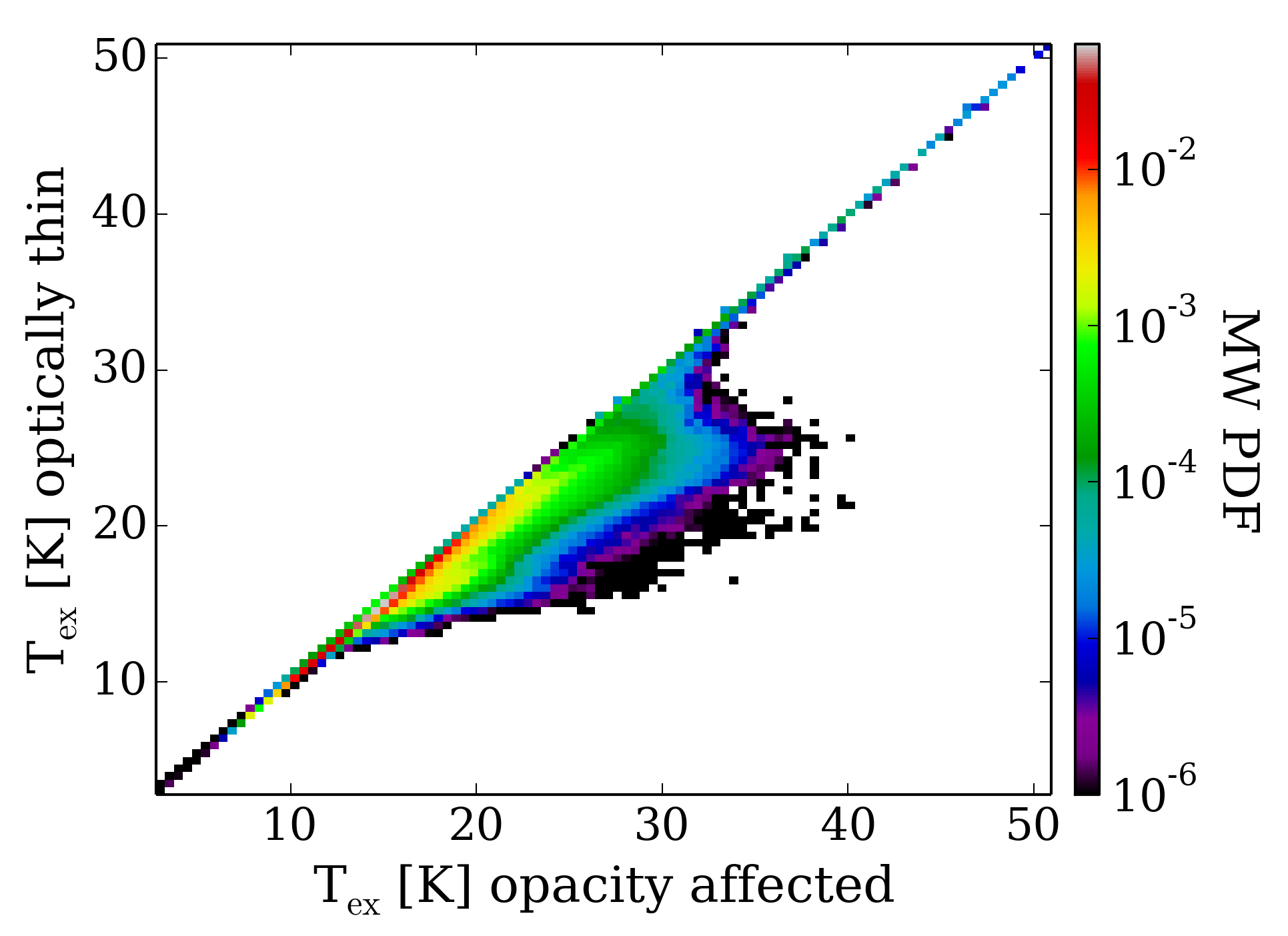}
\caption{Mass weighted 2D-PDF of the excitation temperatures, \Tex, of the opacity affected and the optically thin \CII\ line emission.}
\label{fig.Tex_1213}
\end{figure}

\RADMC\ calculates internally the level populations $x_u$ and $x_l$ for the upper ($u$) and lower level ($l$), respectively, of C$^+$. 
We use this output to recalculate the excitation temperature, \Tex, with 
\begin{equation}
T_{\textrm{ex}} = \frac{- h \nu_{ul}}{k} \times \left[ \textrm{ln}\left( \frac{x_u}{x_l} \frac{g_l}{g_u} \right) \right]^{-1}. 
\end{equation}
Figure~\ref{fig.Tex} shows the 2D-PDF of the gas emissivities as a function of \Tex\ and $T_{\textrm{kin}}$ for the opacity affected (left) and the optically thin (right) \CII\ line emission (bin sizes are $\Delta \textrm{log}_{10}(T_{\textrm{kin}}\;\textrm{[K]}) = 0.07$ and $\Delta \textrm{log}_{10}(T_{\textrm{ex}}\;\textrm{[K]}) = 0.01$). The emissivities were calculated in analogy to Eq.~\eqref{eq.Ltheo}, where we inserted \Tex\ as the temperature. In the low temperature regime ($T \leq 30$~K), LTE conditions are fulfilled ($T_{\textrm{kin}} = T_{\textrm{ex}}$), as the low temperature gas is correlated with high densities in this temperature regime. At $T_{\textrm{kin}} \approx 50$~K, where a large fraction of the emission originates, the excitation temperature becomes significantly smaller than $T_{\textrm{kin}}$ and drops to values around $T_{\textrm{ex}}\approx 20$~K. Therefore the emission is subthermal. The distributions for the opacity affected and optically thin \CII\ line emission appear similar. However, around $T_{\textrm{kin}} \sim 50$~K the excitation temperature tends to higher values for the opacity affected \CII\ line emission compared to the optically thin case. This is caused by the optical depth effects and the resulting radiative excitation. We illustrate the difference between $T_{\textrm{ex}}$ for the opacity affected and the optically thin \CII\ line emission in Fig.~\ref{fig.Tex_1213} with a 2D-PDF (bin size $\Delta T_{\textrm{ex}} = 0.5$~K). \Tex\ deviates most in the optically thin range of 10~K $\lesssim$ \Tex\  $\lesssim$ 35~K. Here the opacity affected \Tex\ is higher by up to 10~K. \\

\begin{figure*}
\centering
\includegraphics[width=\textwidth]{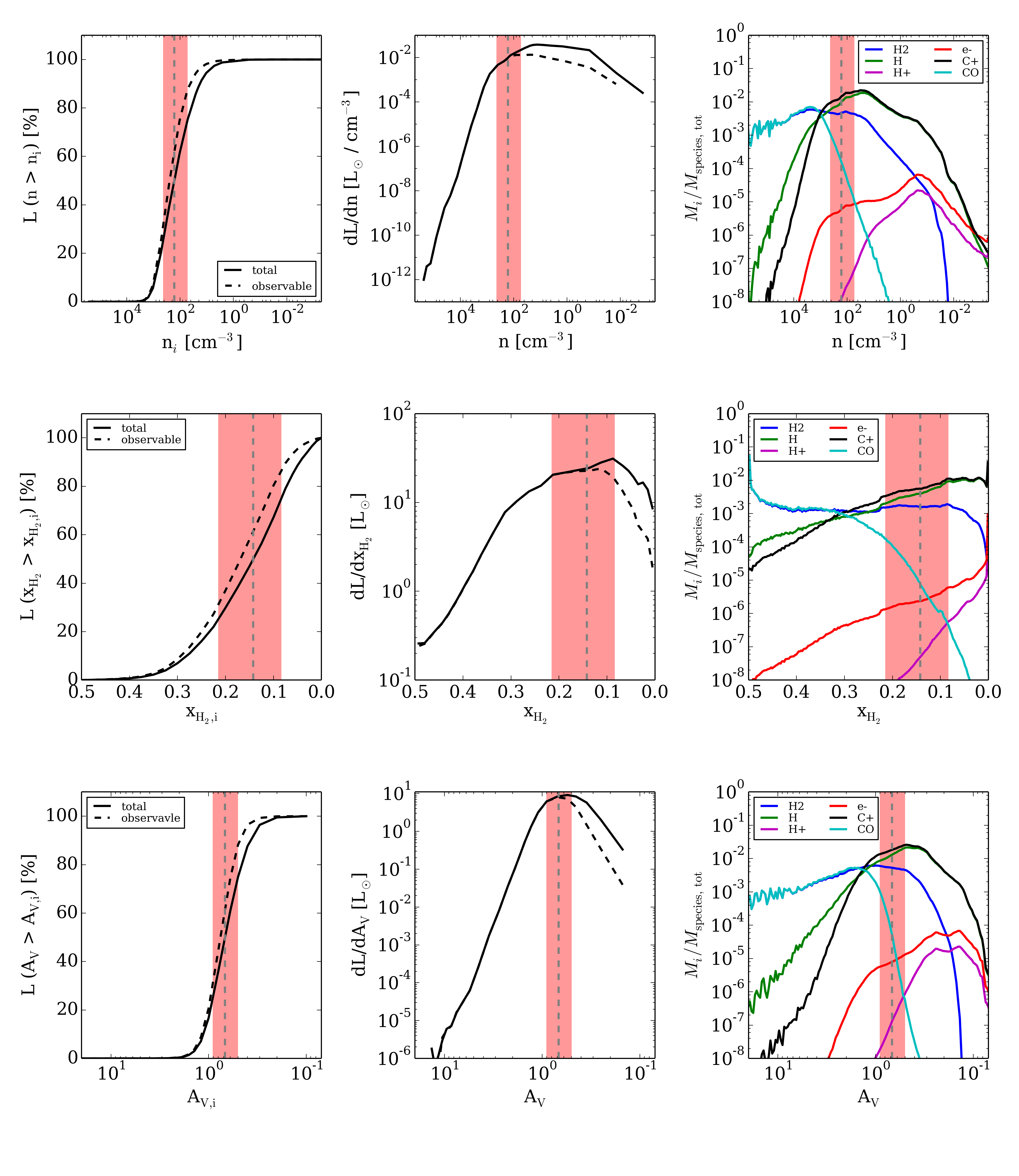}
\caption{Total (solid line) and observable (dashed line) luminosity of the optically thin \CII\ line emission originating from the gas distinguished by its number density with $n > n_i$ (first row), its fractional abundance of H$_2$ ($x_{\textrm{H}_2}$; middle row), and its visual extinction (A$_{\rm V}$; bottom row). In the left column, we show the cumulative distribution and in the middle column its derivative. Note that dense gas appears on the left-hand side of the $x$-axes. The right column depicts the mass-weighted species distributions (see Section~\ref{sec.Analysis_Temp}). The red shaded area and the grey, dashed vertical line mark the median and the interquartile range for the emission of the total map. }
\label{fig.ZoomIn_Origin_07}
\end{figure*}

\subsection{Density dependence}

We further constrain from which density range the \CII\ line emission originates. For this we carry out an analysis in which we only select cells above a limiting number density $n_i$. The left panel of Fig.~\ref{fig.ZoomIn_Origin_07} (first row) shows the resulting cumulative plot, and the middle panel the corresponding derivative. We find the \CII\ line emission to originate from gas with densities between 53~\cmdrei\ and 438~\cmdrei\ (corresponding to the 25$^{\mathrm{th}}$ and the 75$^{\mathrm{th}}$ percentile) with the median of the cumulative luminosity distribution at 166~\cmdrei\ (see also Table~\ref{tab.CumInt}). The same analysis carried out for the optically thick \CII\ line emission results in slightly higher densities (see Table~\ref{tab.CumInt_CIIzwei}). The higher densities are required since part of the emission is absorbed on the way to the observer. The right panel of Fig.~\ref{fig.ZoomIn_Origin_07} (first row) shows the mass-weighted distributions of all species (with bin size $\Delta \textrm{log}_{10}(n\;[\textrm{cm}^{-3}]) = 0.05$). We find the distributions for C$^+$ and atomic hydrogen to be aligned for number densities below $1.2 \times 10^{3}$~\cmdrei, the range from which most of the \CII\ line emission originates. \\

\subsection{Molecular gas dependence}
 
As the \CII\ line emission is used to estimate the fraction of CO-dark H$_2$ \citep{Langer.etal.II.2014}, we investigate how well the \CII\ line emission traces the molecular gas phase in this cloud. We select cells with a particular fractional abundance of molecular hydrogen, $x_{\rm{H}_2}$, where $x_{\rm{H}_2} = 0.5$ if all hydrogen in a cell is is in molecular form. We use the optically thin \CII\ line emission to present the results in the plots. As shown in the left panel of Fig.~\ref{fig.ZoomIn_Origin_07} (second row), we find that the \CII\ line emission starts to increase at $x_{\rm{H}_2,i} \sim 0.4$, where 80\% of the hydrogen is in molecular form. Around 20\% of the emission comes from gas regions with $x_{\rm{H}_{2,i}} \gtrsim 0.23$ ($\sim$47\% of hydrogen in its molecular form). The median value of the cumulative plot (Table~\ref{tab.CumInt}) is at $x_{\rm{H}_2} = 0.14$, meaning that 50\% of the emission comes from gas in which less than 28\% of hydrogen is in molecular form. Thus, we find most of the \CII\ line emission to originate from the atomic gas phase. This is in agreement with the previous results. In the right panel we show the mass-weighted distributions of all species, this time as a function of $x_{\rm{H}_2}$ ($\Delta x_{\rm{H}_2} = 0.02$). C$^+$ is present for all $x_{\rm{H}_2}$, but it only becomes more abundant than CO below $x_{\rm{H}_2} \lesssim 0.3$. This is only slightly deeper in the cloud than $x_{\rm{H}_2} = 0.25$, where the transition from H to H$_2$ dominated gas occurs. \\

\subsection{Dependence on visual extinction}

We further study from which gas phase the \CII\ line emission originates in terms of the visual extinction, A$_{\rm V}$, in the gas. In our version of the \FLASH\ code, A$_{\rm V}$ is calculated for every cell in the computational domain using the \textsc{TreeRay / OpticalDepth} module \citep{Wunsch.etal.2018}. Thus, for any given cell, it is not integrated along a particular line of sight, but instead represents a local weighted average of the values along different lines of sight to that cell \citep[see also][]{Walch.etal.2015}. We calculate the synthetic emission maps including gas with $\textrm{A}_{\textrm{V}}< \textrm{A}_{{\rm V},i}$, and present the results in the bottom row of Fig.~\ref{fig.ZoomIn_Origin_07}. The left panel shows the resulting cumulative plot, its derivative and the mass-weighted distributions of all species ($\Delta \textrm{log}_{10}(\textrm{A}_{\textrm{V}}) = 0.02$). All of the \CII\ line emission comes from gas with $\textrm{A}_{\textrm{V}} < 2$, and 50\% of the emission has $\textrm{A}_{\textrm{V}} < 0.68$. Gas in this regime consists predominantly of atomic hydrogen and ionized carbon, in line with the analyses done before. The median value as well as the 25$^{\mathrm{th}}$ and the 75$^{\mathrm{th}}$ percentile are listed in Table~\ref{tab.CumInt} (and in Table~\ref{tab.CumInt_CIIzwei} for the optically thick case). As done for the other studies we likewise investigate how the result changes when we only take the assumed observable pixels (dashed lines). Since the faint emission stems from gas with low densities, the shielding in this gas is likewise small. Thus, when only taking the observable pixels, we miss the emission from the gas with low A$_{\rm V}$ values, whereas the overall result remains. \\

\section{Correlation with column densities}
\label{sec.XCII}

In extragalactic studies, \CII\ line emission is sometimes used to constrain the mass of the observed system, with the aid of PDR modelling. For example, in studies of submillimetre galaxies \citep{Swinbank.etal.2012} and starburst galaxies \citep{Hailey-Dunsheath.etal.2010} the mass of the molecular and atomic gas, respectively, is calculated directly from the \CII\ line emission. There are likewise studies correlating the \CII\ line emission with column densities. \cite{Goicoechea.etal.2015} study Orion in \CIIzwei\ and \CIIdrei\ and calculate the column density of C$^+$ in the analysed region. In the Milky Way \cite{Langer.etal.II.2014} constrain the C$^+$ column density from the \CII\ line emission, following the work by \cite{Goldsmith.etal.2012}. Inspired by these studies, we analyse the correlation between the \CII\ line emission and the column densities of the total and atomic hydrogen gas, as well as of C$^+$. We note that our molecular cloud is in its formation process before the onset of star formation and therefore no (radiative) feedback processes are considered. Here, we only have a constant ISRF of $G_0 = 1.7$, irradiating the cloud from all directions. Thus, classical PDRs forming around young, massive stars are not considered. However, our simulations are useful for examining \CII\ line emission from gas which is not in a classical PDR. \\

\begin{figure*}
\centering
\includegraphics[width=\textwidth]{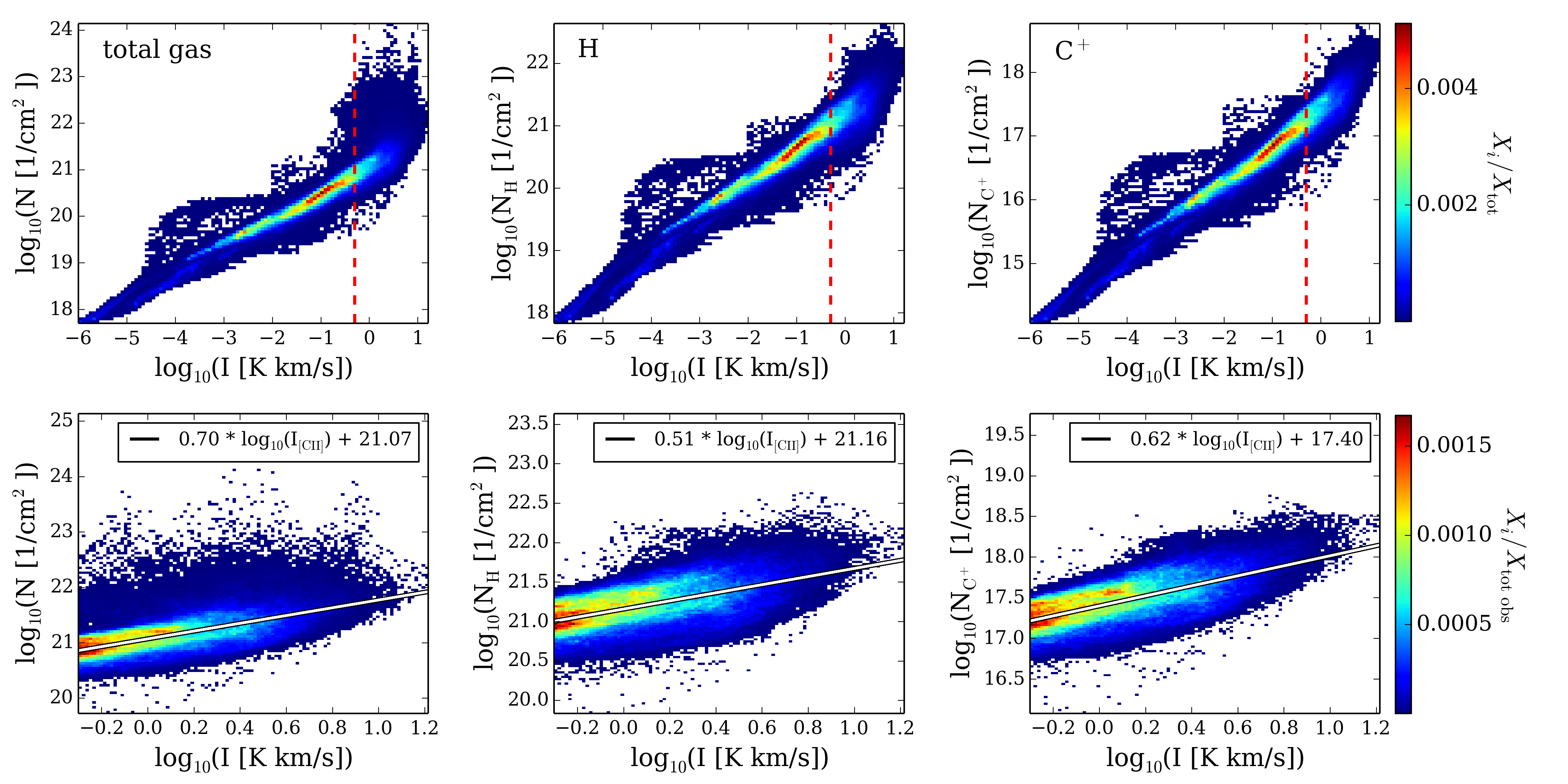}
\caption{Correlation between the integrated \CII\ line intensity (opacity affected) and the column densities of the total gas (left), H (middle) and C$^+$ (right). The top row shows the distribution including the data for the whole map. The red vertical lines mark the assumed observable limit of $I_{[\textrm{CII}]} \geq 0.5$~K~\kms. The bottom row shows only the observable data with their overplotted linear fits. The colour-coding indicates the fraction of pixels with respect to the entire map ($X_i / X_{\textrm{tot}}$) and to the observable area ($X_i / X_{\textrm{tot,~obs}}$). }
\label{fig.INdens}
\end{figure*}

In Fig.~\ref{fig.INdens} we show the correlation between the \CII\ line emission (opacity affected) and the total (left), H (middle), and C$^+$ (right) column densities in the upper row in a 2D-PDF with bin sizes of $\Delta \textrm{log}_{10}(I\;[\textrm{K\,km\,s}^{-1}]) = 0.07$, $\Delta \textrm{log}_{10}(N_i\;[\textrm{cm}^{-2}]) = 0.06$ for the whole map (upper row) and $\Delta \textrm{log}_{10}(I\;[\textrm{K\,km\,s}^{-1}]) = 0.02$, $\Delta \textrm{log}_{10}(N_i\;[\textrm{cm}^{-2}]) = 0.04$ for the observable area (bottom row). The red dashed lines in the plots mark the assumed observable limit at $I_{[\textrm{CII}]} \geq 0.5$~K~\kms. In a double-logarithmic plot the opacity affected \CII\ line emission follows a power law as a function of all three column densities. Higher column densities do not necessarily result in a higher integrated \CII\ line intensity, since the \CII\ line emission is partially optically thick and C$^+$ is converted to CO. Therefore, the distributions steepen in the observable area. In the bottom row of Fig.~\ref{fig.INdens} we show the 2D-PDFs as a function of the opacity affected \CII\ intensity and the column densities, restricted to the observable pixels. The colour-coding indicates the distribution of the fraction of the pixels. The distributions follow power laws that can be fitted\footnote{The fit was obtained using the \textsc{curve\_fit} python package, where the error estimate is based on the Jacobian.} by
\begin{equation}
\label{eq.Ntot}
\begin{split}
\log_{10}\left(\frac{N_{\textrm{tot}}}{\mathrm{cm}^{2}}\right) = (0.702 \pm 0.002) \times \log_{10}\left(\frac{I_{[\textrm{CII}]}}{\mathrm{K\;km\;s}^{-1}}\right) \\
+ (21.0722 \pm 0.0005),
\end{split}
\end{equation}
\begin{equation}
\label{eq.NH}
\begin{split}
\log_{10}\left(\frac{N_{\textrm{H}}}{\mathrm{cm}^{2}}\right) = (0.514 \pm 0.001) \times \log_{10}\left(\frac{I_{[\textrm{CII}]}}{\mathrm{K\;km\;s}^{-1}}\right) \\
+ (21.1595 \pm 0.0004),
\end{split}
\end{equation}
\begin{equation}
\label{eq.NCp}
\begin{split}
\log_{10}\left(\frac{N_{\textrm{C}^+}}{\mathrm{cm}^{2}}\right) = (0.616 \pm 0.001) \times \log_{10}\left(\frac{I_{[\textrm{CII}]}}{\mathrm{K\;km\;s}^{-1}}\right) \\
+ (17.4008 \pm 0.0004). 
\end{split}
\end{equation}

\begin{figure*}
\centering
\includegraphics[width=\textwidth]{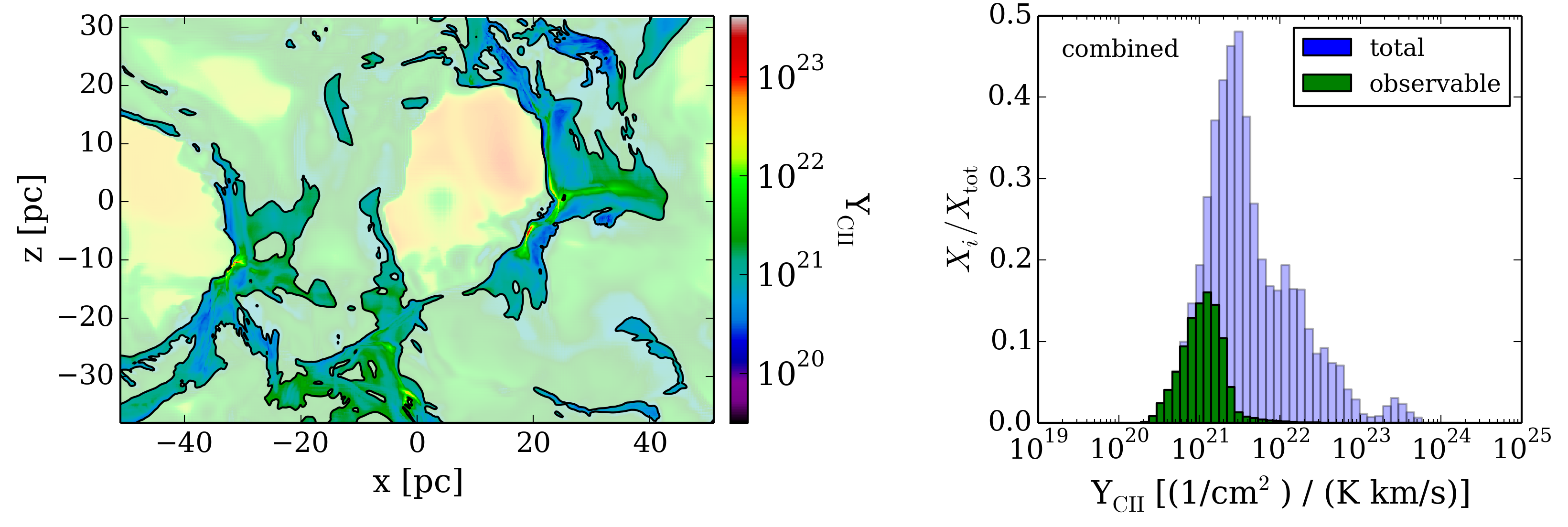}
\caption{Map of \YCII, as defined in Eq.~\eqref{eq.XCII}, for the $x-z$ projection of the L10 simulation (left). The regions within the black contour line in the map belong to pixels that we assume to be observable, fulfilling $I_{[\textrm{CII}]} \geq 0.5$~K~\kms. 
On the right, the normalized histograms of the \YCII\ values for the total map (blue) and the observable pixels (green) for all three projections together are presented. $X_i / X_{\textrm{tot}}$ indicates the fraction of pixels in every bin. }
\label{fig.XCII}
\end{figure*}

\begin{table*}
\centering
\begin{minipage}{\textwidth}
\centering
\caption{List of the 25\%, 50\%, and 75\% values of the \YCII\ factor for the $x-z$ projection and the combined data for all three projections. The upper part of the table contains the \YCII\ values of those regions which are assumed to be observable ($I_{[\textrm{CII}]} \geq 0.5$~K~\kms), and the lower part contains the \YCII\ values obtained when the whole map is taken into account.} 
\label{tab.XCII}
\begin{tabular}{@{}lccccc@{}}
\hline
projection &  & \YCII\ [\IXCII]  \\
 & &\multicolumn{3}{l}{} &  \\
 & & $25$\% & $50$\% (median) & $75$\% & mean  \\
\hline
\hline
$x-z$ & observable & $7.27 \times 10^{20}$ & $1.09 \times 10^{21}$ & $1.53 \times 10^{21}$ & $1.37 \times 10^{21}$\\
combined & observable & $7.48 \times 10^{20}$ & $1.11 \times 10^{21}$ & $1.59 \times 10^{21}$ & $1.44 \times 10^{21}$\\
\hline
$x-z$ & total & $1.65 \times 10^{21}$ & $2.83 \times 10^{21}$ & $5.98 \times 10^{21}$ & $7.23 \times 10^{21}$ \\
combined & total & $1.76 \times 10^{21}$ & $3.21 \times 10^{21}$ & $9.39 \times 10^{21}$ & $1.49 \times 10^{22}$\\
\hline
\end{tabular}
\end{minipage}
\end{table*}
In order to obtain a scale invariant quantity, 
we study the relation between the opacity affected \CII\ integrated intensity ($I$) and the total column density of the molecular cloud ($N_{\rm{tot}}$) with the \YCII\ factor, defined as
\begin{equation}
\label{eq.XCII}
Y_{\textrm{CII}} = \frac{N_{\rm{tot}}}{I}.
\end{equation} 
A constant \YCII\ in the whole map would indicate a linear behaviour between the intensity and the column density. We note that \YCII\ is expected to strongly depend on the local ISRF. The molecular cloud studied here is embedded in an ISRF with $G_0 =1.7$. In clouds which are subject to a weaker/stronger ISRF, \YCII\ might be larger/smaller.\\

In Fig.~\ref{fig.XCII} we plot a map of \YCII\ for the L10 simulation in the $x-z$ projection in the left panel. The regions within the black contour mark the observable region ($I_{[\textrm{CII}]} \geq 0.5$~K~\kms). Over the whole map, the \YCII\ distribution covers a large range with values between 10$^{20} \lesssim$ \YCII\ $\lesssim 10^{24}$~\IXCII. All high values occur only in regions with unobservably low \CII\ intensities. The median of the \YCII\ distribution is at \mbox{$2.83^{+ 3.15}_{- 1.18} \times 10^{21}$~\IXCII\ }, taking all pixels in the map into account, or \mbox{$1.09^{+ 0.44}_{-0.36} \times 10^{21}$~\IXCII\ } if only observable pixels are considered. The other projections behave similarly. We therefore show in the right panel of Fig.~\ref{fig.XCII} the histogram of the fraction of pixels as a function of \YCII\ combined for all three projections ($\Delta \textrm{log}_{10}(Y_{\textrm{CII}}) = 0.08$). The blue coloured histogram represents all values in the map, whereas the green one includes only observable pixels. Table~\ref{tab.XCII} lists the 25$^{\textrm{th}}$ and 75$^{\textrm{th}}$ percentiles as well as the median value of the \YCII\ factors for the $x-z$ projection and the data combined for all three projections. If we assume that the observation of a molecular cloud is unresolved, we would obtain a single intensity value for the whole region ($\sim$0.4~K~\kms\ as the average integrated intensity over the whole map), and might have one single value for the column density ($\sim 6.3 \times 10^{20}$~cm$^{-2}$). Using these simplified values for obtaining \YCII, we find $\sim1.5 \times 10^{21}$~\IXCII. \\

We further analyse the correlation between \YCII\ and $I_{[\textrm{CII}]}$ for the observable part in more detail. Expressing Eq.~\eqref{eq.Ntot} in terms of \YCII\ yields 
\begin{equation}
\label{eq.XCII_fit}
\begin{split}
\log_{10}&\left(\frac{Y_{\textrm{CII}}}{\textrm{cm}^{-2} (\textrm{K km s}^{-1})^{-1}}\right) =   \\ - (0.298 &\pm 0.165) \times \log_{10}\left(\frac{I_{[\textrm{CII}]}}{\textrm{K km s}^{-1}}\right) 
+ (21.0722 \pm 0.0005). 
\end{split}
\end{equation}
The $y$-intercept of this relation ($y = 21.0722 \pm 0.0005$) corresponds to a column density of $N_{\mathrm{tot}} = 10^y \approx 10^{21}$~cm$^{-2}$. This is approximately the column density at which $A_{\rm V} = 1$, and where the transition between atomic to molecular gas happens. Taking Eq.~\eqref{eq.XCII_fit} might be better for extracting the \YCII\ value and the total column density $N_{\mathrm{tot}}$ for an observation than just taking e.g. the median value of \YCII. To prove this, we calculate the distribution of \YCII\ around the fitting function  Eq.~\eqref{eq.XCII_fit} (not shown). The width of this distribution, measured with the interquartile range, $d_{\textrm{iqr}} = 7.5 \times 10^{20}$~\IXCII, is smaller compared to the distribution of the observable \YCII\ values shown in the right panel of Fig.~\ref{fig.XCII}, with $d_{\textrm{iqr}} = 8.4 \times 10^{20}$~\IXCII, which represents a small improvement. To further estimate the difference between the fit and the \YCII\ distribution from the simulation, we calculate the deviation between the distribution of \YCII\ values and the ones calculated with Eq.~\eqref{eq.XCII_fit} as 
\begin{equation}
\Delta_{Y_{\textrm{CII}}} = |Y_{\textrm{CII}}^{\textrm{fit}} - Y_{\textrm{CII}}^{\textrm{map}}| \times \frac{100\%}{Y_{\textrm{CII}}^{\textrm{fit}}}.
\end{equation}
We find that 25\%, 50\%, 75\% of the pixels have $\Delta_{Y_{\textrm{CII}}} \approx 16$\%, 33\%, 54\%, respectively. \\

\begin{figure}
\centering
\includegraphics[width=80mm]{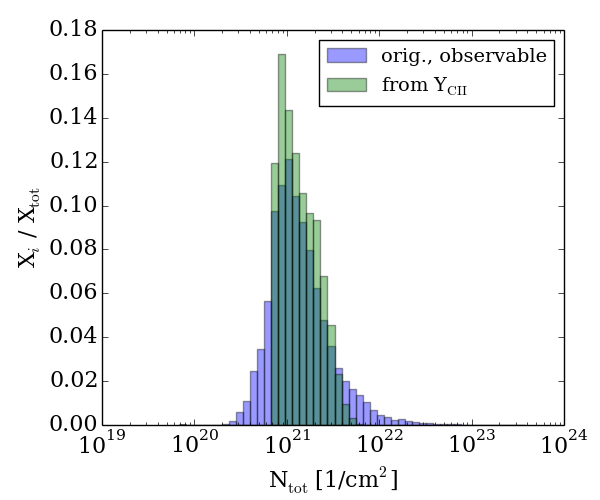}
\caption{Histogram of the column densities of the observed area (blue), and of the column densities calculated with Eq.~\eqref{eq.XCII_fit} using the opacity affected \CII\ integrated intensity. $X_i / X_{\textrm{tot}}$ indicates the fraction of pixels as a function of the logarithmically binned column densities $N_{\rm{tot}}$. We underestimate the projected mass by 22\% to 35\%.}
\label{fig.TestXCII}
\end{figure}

We test the practical value of the \YCII\ factor fitting by applying Eq.~\eqref{eq.XCII_fit} to the simulation L10 in the $x-z$ projection. The whole simulation box contains a total mass of $8.4 \times 10^4$~M$_{\odot}$. About 70\% of this mass ($m_{\textrm{proj}} = 5.8 \times 10^4$~M$_{\odot}$) is situated in regions of the $x-z$ projection above the assumed detection limit ($I_{[\textrm{CII}]} \geq 0.5$~K~\kms). If we take only the pixels above the detection limit and recalculate the total mass of the cloud using Eq.~\eqref{eq.XCII_fit}, we obtain $m_{\textrm{cloud, YCII}} = (4.1^{+ 0.405}_{- 0.320}) \times 10^4$~M$_{\odot}$, which is in the range of 65\% to 78\% of the projected mass. In Fig.~\ref{fig.TestXCII} we show the histogram of the column densities of the observed area (blue) and of the column densities recalculated from the integrated opacity affected \CII\ intensity (green) with a bin size of $\Delta \textrm{log}_{10}(Y_{\textrm{CII}}) = 0.08$. We do not reproduce the small column densities at low intensities and likewise miss the high column density regions. \\

However, when using the mean value and the interquartile range of the \YCII\ distribution (\mbox{$1.11^{+ 0.48}_{- 0.37} \times 10^{21}$~\IXCII}) for estimating the mass, we obtain $m_{\textrm{cloud, YCII}} = (11.25^{+ 4.85}_{- 3.65}) \times 10^4$~M$_{\odot}$. This corresponds to masses between 130\% and 280\% of $m_{\textrm{proj}}$. We therefore conclude that \YCII\ is not constant, but it is better estimated when assuming that it scales as $I_{[\textrm{CII}]}^{-0.3}$ (Eq. ~\eqref{eq.XCII_fit}). \\

\section{Discussion}
\label{sec.Discussion}

\subsection{Discussion of general results}

In observations, optical depth effects of the \CII\ line emission are mainly discussed in the context of regions with star formation feedback, as in classical PDRs \citep{Boreiko.Betz.1997, Stacey.etal.2010, Graf.etal.2012, Ossenkopf.etal.2013, Neri.etal.2014, Gullberg.etal.2015}. When studying molecular clouds in observations, the \CII\ line emission is assumed to be optically thin \citep[e.g.][]{Langer.etal.2010, Pineda.etal.2013, Langer.etal.I.2014}. In our model of a forming molecular cloud before the onset of star formation we find the \CII\ line emission to be optically thick in $\sim$40\% of the assumed observable region (defined as the area where $I_{[\textrm{CII}]} \geq 0.5$~K~\kms) with optical depths up to $\tau \sim 10$. However, only a small part of the map is affected by high optical depths. Therefore, the approach of the \CII\ line emission being ``effectively optically thin'' holds for most of the emission \citep{Goldsmith.etal.2012}, so that the \CII\ line emission is linearly correlated with the C$^+$ column density (c.f. Fig.~\ref{fig.INdens}). \\

For studying the local properties (density, temperature, visual extinction, and molecular gas fraction) of the \CII\ emitting gas, we perform a second set of radiative transfer simulations, where we treat the C$^+$ gas as optically thin. For this we reduce the C$^+$ abundance such that this corresponds to emission from $^{13}$C$^+$. We find that in the studied young molecular cloud the \CII\ line emission originates from the cold, atomic gas phase with temperatures between $\sim$40~K  and $\sim$65~K and number densities of the total gas between $\sim$50~\cmdrei\ and $\sim$440~\cmdrei. In the absence of turbulence, molecular hydrogen forms in the ISM on a timescale $t_{\textrm{H}_2} \approx 1\,\textrm{Gyr} / n$ \citep{Hollenbach.etal.1971}, corresponding to a timescale of between 2 and 20~Myr for gas in this density range. The presence of turbulence can accelerate the H$_{2}$ formation rate by a factor of a few \citep{Glover.etal.2010}, but even when we account for this, much of the \CII-bright gas still has an H$_{2}$ formation timescale that is comparable to or longer than the age of the cloud at the moment that we analyze it. It is therefore quite likely that the atomic hydrogen in the cloud is still in the process of becoming molecular at the time at which we analyse the cloud (see Section~\ref{sec.zoom} above). During this transition phase, we expect much of the \CII\ emission to come from gas dominated by atomic hydrogen, in line with what we find in this study, but in more evolved clouds it is likely that a greater fraction of the emission would come from H$_{2}$-dominated gas (c.f.\ \citealt{Glover.Smith.2016, Clark.etal.2018}). \\

As our chemical network includes carbon as CO and C$^+$, but misses the atomic form, we might overestimate the amount of CO in the simulation \citep{Glover.Clark.2012}. Including the atomic carbon would also reduce the amount of C$^+$. However, the main effect of this would be to shift the ranges in temperature and density of the \CII\ emitting gas to slightly higher temperatures and lower densities. Therefore, the correlation between \CII\ emission and the atomic gas in the transition phase would remain. \\

\subsection{Comparison with observations}
There are studies analysing the origin of the emission on a Galaxy-wide scale, as it is done in the \mbox{GOT C+ survey} along different lines of sight \citep[e.g.][]{Pineda.etal.2013, Velusamy.Langer.2014}. By comparing the \CII\ components with ancillary \HI, $^{12}$CO, $^{13}$CO and C$^{18}$O emission, they identify the phase from which the \CII\ line emission originates. According to \cite{Pineda.etal.2013} and \cite{Pineda.etal.2014}, 30--50\% of the \CII\ line emission stems from PDRs, 25--28\% from CO-dark~H$_2$, 21--25\% from gas associated with \HI, and 4--20\% from the ionized gas phase. Since our simulations do not include the feedback from stars, we only account for the 50\% to 70\% of the emission that does not come from PDRs or the ionized gas phase. We find that about 20\% of the \CII\ line emission stems from H$_2$ dominated gas, where more than 50\% of hydrogen is in form of H$_2$, and 80\% of the \CII\ line emission originates from the atomic gas phase. Thus, our fraction associated with the atomic gas phase is larger than found by \cite{Pineda.etal.2013, Pineda.etal.2014}. The difference likely arises from the fact that the GOT C+ survey includes numerous molecular clouds, with a range of ages. Those clouds which are young and are forming, such as the cloud in our simulation, will predominately contribute to the \CII\ emission associated with the atomic phase. Clouds which are more evolved will predominately contribute to the \CII\ emission associated with the molecular phase. We will investigate the effect of the cloud age on the \CII\ emission in a follow-up paper. \\

The \CII\ line emission is studied in infrared dark clouds e.g. by \cite{Beuther.etal.2014}, or recently \cite{Bisbas.etal.2018}. They find the \CII\ line emission to exhibit diverse morphologies, ranging from a non-detection in the most quiescent region, to stronger emission in more actively star-forming regions.  According to \cite{Beuther.etal.2014}, the \CII\ line emission traces the environment around the dense gas and is sensitive to the external UV field. Our simulations confirm that the \CII\ line emission is a good tracer for the envelope of the densest parts of molecular clouds, tracing mostly atomic hydrogen. We further suggest that the \CII\ line emission is even detectable before the onset of star formation. However, the emission properties as well as the derived \YCII\ (see Section~\ref{sec.XCII}) are expected to depend on the local interstellar radiation field. Here we study a cloud which is irradiated by an ISRF of $G_0=1.7$. For a UV dominated cloud, we expect a correlation with the molecular gas, observable in CO. Thus, it is therefore important to have multispecies observations to better understand the physical properties of a cloud. \cite{Beuther.etal.2014} further find promising kinematic signatures in their \CII\ detections, and a strong velocity gradient for G48.66. From this, they suggest to discern between various cloud assembly processes by using observations of \CII. In our model of a forming molecular cloud we find the \CII\ emission to stem from the atomic envelope. Since the cloud is still in formation and the gas collapsing inwards, we expect this movement of the gas to be visible in kinematic signatures of the \CII\ line. A future work will address these kinematic signatures.

\subsection{Comparison with simulations}
The origin of the \CII\ line emission is likewise studied in simulations. 
\cite{Smith.etal.2014} simulate the ISM within a torus mimicking the outer parts of the Milky Way. Their simulations are done with the moving mesh code \AREPO\ \citep{Springel.2010}, accounting for a galactic potential, but neglecting the feedback from star formation and neglecting self-gravity. They simulate the chemical evolution of the gas with the same chemical network as we use  \citep{Glover.MacLow.2007.I, Glover.MacLow.2007.II, Nelson.Langer.1997}. In \cite{Glover.Smith.2016} they study the \CII\ line properties in those simulations. Instead of taking a radiative transfer code for synthetic observations, they estimate the \CII\ emissivity in every cell assuming that the line is optically thin. Thus, the emission they calculate is comparable to the optically thin \CII\ line emission in our work. \cite{Glover.Smith.2016} find a small part of the \CII\ line emission to originate from the warm neutral medium (WNM) with densities around 0.2~\cmdrei\ and temperatures around $10^{4}$~K. We do not see a contribution from the WNM. However, our zoom-in region only contains the molecular cloud, and we therefore do not account for the large volume of warm gas that \cite{Glover.Smith.2016} have in their simulation. They obtain a larger contribution from the cold atomic, and CO-dark molecular gas, which is broadly distributed around densities of 20~\cmdrei\ and temperatures of 100~K. Similar to \cite{Smith.etal.2014} and \cite{Glover.Smith.2016}, we do not include stellar feedback in the underlying 3D simulation, and therefore only study the \CII\ line emission from the cold, molecular and atomic gas, but not from PDRs. Our results show in the same way a large contribution from the cold gas phase, although we find that most of the emission comes from gas in the temperature range $40 < T < 65$~K, while \cite{Glover.Smith.2016} find that the largest contribution comes from gas close to 100~K. \cite{Glover.Smith.2016} further distinguish the gas by its amount of atomic hydrogen and find that half of the emission from the cold gas phase is associated with H$_{2}$-dominated regions (defined as having more than 50\% of the hydrogen mass in the form of H$_2$), counting this fraction as CO-dark H$_2$. We, on the other hand, only find $\sim$20\% of the \CII\ line emission associated with the molecular gas phase following the same definition. These differences may be due to the fact that the simulation presented in \cite{Glover.Smith.2016}  does not contain self-gravity, thus the clouds extend to much lower densities and are unlikely to ever form much CO. Here, we focus on a single gravitationally-bound molecular cloud. \\

\cite{Accurso.etal.2017} investigate in simulations how much of the \CII\ line emission ($f_{[\textrm{CII}],\textrm{mol}}$) is correlated with the molecular gas phase. Their setup consists of a multiphase 3D radiative interface that couples the stellar spectrophotometric code \textsc{starburst99} \citep[][]{Leitherer.etal.1999, Leitherer.etal.2010, Vazquez.Leitherer.2005, Conroy.2013} with the photoionization and astrochemistry codes \textsc{mocassin} \citep{Ercolano.etal.2003} and \textsc{3d-pdr} \citep{Bisbas.etal.2012}. They model entire star-forming regions and thus, their simulations include the part of feedback that we do not account for. On the other hand, when modelling the spherically symmetric geometry with an ionizing source in the centre of the cloud, hydrodynamical effects, as turbulence and shocks, are neglected \citep{Accurso.etal.2017}. Our simulations in contrast focus on the hydrodynamical evolution of the gas before the onset of star formation in a turbulent molecular cloud. \cite{Accurso.etal.2017} assume that the physical conditions found in each of their clouds can represent the average physical conditions found on galaxy-wide scales. By using a Bayesian formalism and statistical weights for the clouds, they fit $f_{[\textrm{CII}],\textrm{mol}}$ as a function of the density, dust mass fraction, specific star formation rate (SSFR), and the metallicity. In general, $f_{[\textrm{CII}],\textrm{mol}}$ increases with the density and decreases with SSFR. For a galaxy matching the parameter space of the Milky Way, they find that around 75\% of the \CII\ emission is correlated with the molecular gas (where the molecular gas is defined as all gas with more than 1\% of hydrogen in the form of H$_2$). In our observable area, 99.8\% of the \CII\ emission comes from molecular gas defined in this way. Our simulations \citep[][S17]{Walch.etal.2015, Girichidis.etal.2016} represent an ISM, in which C$^+$ is formed dynamically, and turbulent mixing of the chemical species is taken into account (see S17). The model of \cite{Accurso.etal.2017}, on the other hand, represents an ISM dominated by UV radiation. They further assume that the chemistry has reached equilibrium, which is both not required and not the case in our setup. The different approaches have an impact on the formation of the PDRs in the model, that in turn cause the difference in the results. \\

Overall, we find that the substructure of the molecular cloud, which is set by gas accretion, turbulence, self-gravity, and the local radiation field, is essentially shaping the \CII\ emitting regions, which envelop the dense, CO-rich filaments and cores. The origin of the \CII\ emission is hence sensitive to the physical conditions as well as the evolutionary stage of the molecular cloud. \\

\section{Conclusion}
\label{sec.Conclusion}

We present synthetic \CII\ line emission maps of a molecular cloud embedded in a piece of a galactic disc. The turbulent molecular cloud simulation has been carried out with the 3D AMR code \FLASH\ \citep{Fryxell.2000, Dubey.etal.2008} including a simple chemical network and is presented in the SILCC-Zoom project \citep{Seifried.etal.2017}. We study this molecular cloud at an early evolutionary stage, before the onset of star formation, with an ISRF of $G_0 = 1.7$. The radiative transfer simulations for the \CII\ line are done with \RADMC\ \citep{Dullemond.2012.II}. We do not assume the gas to be in LTE. Instead, we calculate the level populations by considering collisions with H$_2$, H, and electrons and use the LVG approximation.\\

We investigate the influence of the spatial resolution on the synthetic emission maps. Different versions of the zoom-ins were done for spatial resolutions ranging from d$x = 3.9$~pc to d$x = 0.122$~pc. We find the \CII\ intensity distributions of the maps to converge for resolutions better than $\sim$0.25~pc. All further analyses are based on the highest resolution run with d$x = 0.122$~pc (run L10). \\

If we assume an observable limit of $I_{[\textrm{CII}]} \geq 0.5$~K~\kms, then we recover 80\% of the total \CII\ luminosity. This demonstrates that a molecular cloud should be observable in \CII\ line emission prior to the onset of massive star formation, potentially allowing us to trace the assembly of the cloud. The observable luminosity stems from 16\% -- 27\% of the total area of the map. \CIIzwei\ becomes optically thick in $\sim$10\% of the total map area, corresponding to $\sim$40\% of the observable area. Although the \CII\ line emission behaves as ``effectively optically thin'' \citep{Goldsmith.etal.2012}, observers should bare in mind that the observed \CII\ emission becomes optically thick up to $\tau \sim 10$. \\

To determine the physical properties of the \CII\ emitting gas, we compare the opacity affected \CIIzwei\ emission with an optically thin equivalent, corresponding to \CIIdrei\ emission. We do this by reducing the number density of C$^+$ by a factor of 107. We find the \CII\ line emission to be emitted from gas in the temperature range of 43~K and 64~K and in the number density range of 53~cm$^{-3}$ and 438~cm$^{-3}$, while the visual extinction is between 0.50 and 0.91. We further study the correlation with molecular hydrogen and find that $\lesssim$20\% of the \CII\ emission comes from gas which is dominated by H$_2$, meaning that more than 50\% of the hydrogen is in the form of H$_2$. Thus, we conclude that the \CII\ line emission is generally correlated with atomic hydrogen in transition to the molecular phase in our young, dynamically evolving molecular cloud. As a consequence, we conclude that \CII\ is not a suitable tracer for CO--dark H$_2$ in this scenario. Instead, we expect the \HI\ line to be correlated with the \CII\ line emission. \\

In observational studies the \CII\ line emission is used to constrain the total column density as well as the C$^+$ column densities \citep{Langer.etal.II.2014, Goicoechea.etal.2015}. Here we define a new quantity, the \YCII\ factor, as the ratio between the total gas column density and the integrated \CII\ line emission, that can be used in those studies. The median value is \mbox{\YCII\ $\approx 1.1 \times 10^{21}$~\IXCII}, but to constrain the total gas column density (and hence the total mass of the cloud) using \YCII\ and $I_{[\textrm{CII}]}$, it is better to use \YCII\ $\propto$ $I_{[\textrm{CII}]}^{-0.3}$. We note that this particular value of \YCII\ applies to our young molecular cloud, which is embedded in a uniform ISRF with strength $G_0=1.7$. We expect \YCII\ to be lower / higher in regions with a higher / lower ISRF.

\section*{Acknowledgments}
We thank the anonymous referee for their comments, that were very helpful to improve the work. We are grateful to the support of the Deutsche Forschungsgemeinschaft (DFG) Priority Programme 1573 ``The Physics of the ISM'', which funded large parts of this collaborative project. Furthermore, AF, SW, DS, and VOO acknowledge the DFG Collaborative Research Center 956 (Sonderforschungsbereich SFB 956) ``Conditions and impact of star formation'' and the Bonn--Cologne Graduate School for Physics and Astronomy (BCGS) for their financial support. SW and SC thank the European Research Council for funding through the ERC Starting Grant ``RADFEEDBACK'' no. 679852. RW acknowledges support from Albert Einstein center for gravitation and astrophysics, Czech Science Foundation grant 14-37086G and by the institutional project RVO: 67985815. 
RSK and SCOG acknowledge support from the DFG via SFB 881 ``The Milky Way System'' (sub-projects B1, B2 and B8). 
RSK acknowledges support from the European Research Council under the European Community's Seventh Framework Programme (FP7/2007-2013) via the ERC Advanced Grant STARLIGHT (project number 339177). 
PG further acknowledges the funding from the European Research Council under ERC-CoG grant CRAGSMAN-646955.
We further gratefully acknowledge the Gauss Centre for Supercomputing e.V. 
(\url{www.gauss-centre.eu}) for funding this project by providing computing time on the GCS Supercomputer SuperMUC at Leibniz Supercomputing Centre (\ul{www.lrz.de}). AF likes to thank the Regional Computing Center of the University of Cologne (RRZK) for providing computing time on the DFG-funded High Performance Computing (HPC) system CHEOPS. Furthermore, AF thanks RW, Franti\v{s}ek Dinnbier and Tereza Je\v{r}\'{a}bkov\'{a} for the python package reading the SILCC data. 
The software used to carry out the original simulations was
in part developed by the DOE NNSA-ASC OASCR Flash Center at the
University of Chicago. We thank M. Turk and the \texttt{yt} community for the
\texttt{yt} project \citep{Turk.etal.2011}.

\appendix

\section{Details of the radiative transfer simulations}
\label{app.RADMC}

\subsection{\RADMC}
\RADMC\ provides the possibility of calculating the line emission of ions under the condition that the local thermodynamic equilibrium is not fulfilled. 
It solves the radiative transfer equation
\begin{equation}
\frac{dI_{\nu}(\zeta, s)}{ds} = j_{\nu}(\zeta, s) - \alpha_{\nu}(\zeta, s) I_{\nu} (\zeta, s),
\end{equation}
with the emission and absorption coefficients $ j_{\nu}(\zeta, s)$ and  $\alpha_{\nu}(\zeta, s)$. 
Following the local notation of \cite{vanderTak.etal.2007}, these coefficients are calculated internally in \RADMC\ by
\begin{equation}
j_{\nu}(\zeta, s) = \frac{h \nu}{4 \pi} n_{\textrm{C}^+} x_u A_{ul} \phi_{ul}(\zeta, s)
\end{equation}
\begin{equation}
\alpha_{\nu}(\zeta, s)= \frac{h \nu}{4 \pi} n_{\textrm{C}^+} (x_l B_{lu} - x_u B_{ul}) \phi_{ul}(\zeta, s).
\end{equation}
Here, $n_{\textrm{C}^+}$ is the number density of C$^+$, $x_u$ and $x_l$ are the fractional level populations of the upper ($^2$P$_{3/2}$, indicated with $u$) and lower ($^2$P$_{1/2}$, indicated with $l$) level. $A_{ul}$, $B_{ul}$ and $B_{lu}$ are the Einstein coefficients. The profile function $\phi_{ul}(\zeta, s)$ is used in its co-moving form $\tilde{\phi}_{ul}$ around the line-center frequency $\nu_{ul}$. It is approximated by a Doppler profile: 
\begin{equation}
\tilde{\phi}_{ul} = \frac{c}{a_{\textrm{tot}}\nu_{ul}\sqrt{\pi}}\exp\left\{-\frac{c^2(\nu-\nu_{ul})^2}{a^2_{\textrm{tot}}\nu^2_{ul}}\right\},
\end{equation}
where $a_{\textrm{tot}} = \sqrt{a_{\textrm{therm}}^2 + a_{\textrm{turb}}^2}$ is the line-width composed of the contributions from the thermal line-width, $a_{\textrm{therm}} = \sqrt{\frac{2 k_{\textrm{B}} T}{\mu m_{_{\rm H}}}}$, and a turbulent line-width $a_{\textrm{turb}}$. We set $a_{\textrm{turb}} =  a_{\textrm{therm}}$. The kinetic temperature is inserted for $T$, $k_{\textrm{B}}$ is the Boltzmann constant, $m_{_{\rm H}}$ is the mass of a hydrogen atom, and $\mu$ is the molecular weight of the emitting $^{12}$C$^+$ particle ($\mu = 12$). \\

To calculate the emission and absorption coefficients, and hence, the radiative transfer equation, the fractional level populations of C$^+$ ($x_u$, $x_l$) is needed. The level populations of C$^+$ is considered not to be in the local thermal equilibrium (LTE), since a large fraction of the gas is in dilute medium. \RADMC\ uses the  large velocity gradient \citep[LVG,][]{Sobolev.1957, Ossenkopf.1997, Shetty.etal.2011.I} approximation, also known as Sobolev approximation, for calculating the level populations. In that approximation the optical depth $\tau_{ul}^{\textrm{LVG}}$ is derived from the absolute value of the velocity gradient $\left| \nabla \vec{v} \right|$. In general, the optical depth is defined as d$\tau_{ul} \equiv \alpha_{ul} \textrm{d}s$. Following the local formulation of \cite{vanderTak.etal.2007}, this can be expressed as
\begin{equation}
\label{eq.tau}
\tau_{ul}^{\textrm{LVG}} = \frac{ch}{4 \pi} \frac{n_{\textrm{C}^+}}{1.064 \left| \nabla \vec{v} \right|} \left(x_l B_{lu} - x_u B_{ul} \right). 
\end{equation}
This is used to calculate the escape probability $\beta_{ul}$. For the line within the LVG approximation the escape probability is set to  
\begin{equation}
\label{eq.beta}
\beta_{ul} = \frac{1 - e^{-\tau_{ul}}}{\tau_{ul}}, 
\end{equation}
\citep{Sobolev.1957, vanderTak.etal.2007}. Although this is an assumption for a different geometry, it is used in \RADMC\ \citep[e.g.][for a further discussion]{Ossenkopf.1997}. $\beta_{ul}$ is in turn taken to derive the line integrated mean intensity $J_{ul}$ as
\begin{equation}
J_{ul} = (1 - \beta_{ul}) S_{ul} + \beta_{ul} J_{ul}^{\textrm{bg}}. 
\end{equation}
$S_{ul}$ denotes the source function with 
\begin{equation}
S_{ul} = \frac{j_{\nu}}{\alpha_{\nu}}, 
\end{equation}
and $J_{ul}^{\textrm{bg}}$ is the background radiation field at the rest frequency. Since at $\nu_{[\textrm{CII}]}$ the continuum background radiation in the diffuse interstellar medium is low \citep{Draine.2011} and in observations the continuum background is likewise subtracted, we neglect it for our purpose. \\

\begin{figure}
\centering
\includegraphics[width=80mm]{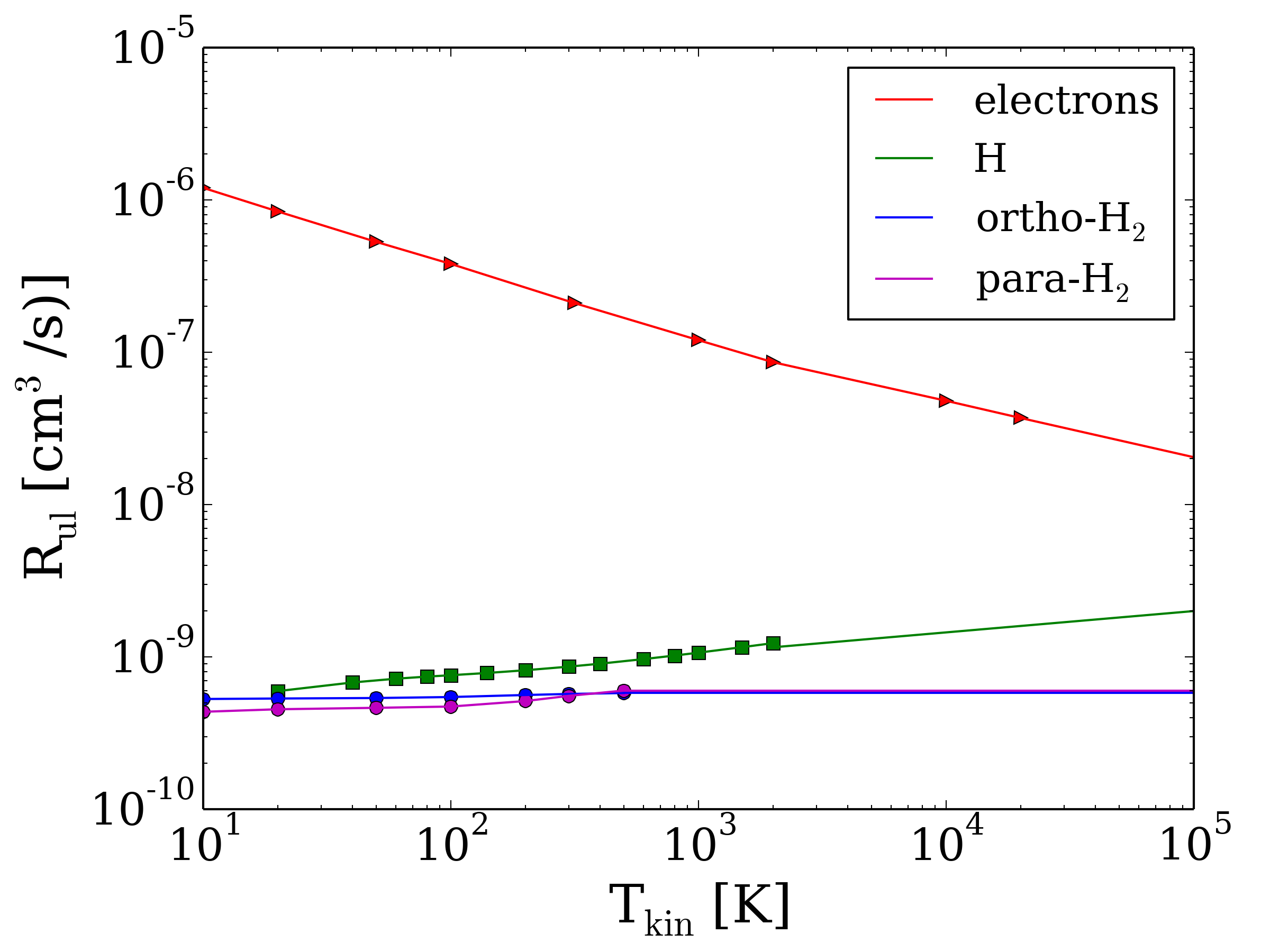}
\caption{The collisional de-excitation rates for the different collisional partners of C$^+$ as a function of the kinetic temperature T$_{\textrm{kin}}$. The rates for the electrons, atomic hydrogen and ortho- and para-molecular hydrogen (red triangles, green squares and blue and magenta circles, respectively) are taken from the Leiden database \citep{Schoeier.etal.2005}. 
To cover the full temperature range of the simulations, we interpolate these database rates and extrapolate them towards higher temperatures. 
For both ortho- and para-H$_2$ we assume the database value at $T = 500$~K for higher temperatures.}
 \label{fig.rates}
\end{figure}

A further ingredient needed for calculating the level populations is the set of collisional excitation and de-excitation rates between C$^{+}$ and its main collision partners, ortho-H$_2$, para-H$_2$, H and e$^-$. We use collisional de-excitation rates taken from the Leiden Atomic and Molecular Database \citep[LAMDA,][]{Schoeier.etal.2005}. For ortho-H$_{2}$ and para-H$_{2}$, these come originally from \citet{Wiesenfeld.Goldsmith.2014}, for atomic H from \citet{Barinovs.etal.2005} and for electrons from \citet{Wilson.Bell.2002}. The fits given in LAMDA do not cover the whole range of temperatures encountered in our simulation, and so at high temperatures we extrapolate the rates following \cite[][ see Fig.~\ref{fig.rates}]{Goldsmith.etal.2012}. Given the collisional de-excitation rate coefficient for each collision partner, the total rate then follows as
\begin{equation}
\label{eq.Cul}
C_{ul} = \sum_{\textrm{cp}} n_{\textrm{cp}} R_{\textrm{ul}}^{\textrm{cp}},
\end{equation}
where $n_{\rm cp}$ is the number density of the collision partner. The collisional excitation rate then follows via the principle of detailed balance:
\begin{equation}
C_{lu} = C_{ul}  \frac{g_{u}}{g_{l}} e^{- \frac{\Delta E}{k_B T}},
\end{equation}
where $g_{u}$ and $g_{l}$ are the statistical weights of the upper and lower levels respectively.\\

The Einstein coefficients and the collisional rates per C$^+$ ion are used to calculate the level populations of the upper and lower level ($x_u$ and $x_l$, respectively).They are derived by solving the equations of statistical equilibrium. In the case of the fine structure transition of the C$^+$ ion with the two levels $u$ and $l$ this reduces to 
\begin{equation}
\label{eq.stateq}
- \left[ x_u A_{ul} \beta_{ul} + (x_u B_{ul} - x_l B_{lu}) \beta_{ul} J_{ul}^{\textrm{bg}} \right] + \left[x_l C_{lu} - x_u C_{ul} \right]  = 0.
\end{equation} 
Since the escape probability $\beta_{ul}$ changes with the level populations, equations \eqref{eq.beta} and \eqref{eq.stateq} are iteratively solved until convergence is reached. \\

In cases where the co-moving line-width is narrower than the Doppler shift, there could be velocity channels with no contribution to the emission in the simulations. However, these so-called Doppler jumps are numerical artefacts and do not occur in nature. In order to avoid them in our simulations we use the Doppler-catching method, described by \cite{Pontoppidan.etal.2009}, which ensures a smooth behaviour \citep[see][]{Shetty.etal.2011.I, Shetty.etal.2011.II}. \\

\subsection{Calculating the collisional rates}

\noindent{\sc $R(\textrm{H}_2)$:} 
For molecular hydrogen the LAMDA database provides de-excitation rate coefficients for temperatures \mbox{$T_{\textrm{kin}} \leq 500$~K} for ortho- and \mbox{para-H$_2$} \citep{Wiesenfeld.Goldsmith.2014}. There is no good fit for the coefficients distinguishing between the two nuclear spin states that could be extrapolated to higher temperatures. We expect that collisions between C$^+$ and H$_2$ at temperatures $T > 500$~K will not contribute significantly to the \CII\ line emission, as in the simulations there is a negligible amount of H$_2$ in regions warmer than 500~K ($\sim$0.1\% of the H$_2$ mass). Therefore, at $T > 500$~K, we assume that the collisional rate coefficients for collisions with ortho- and para-H$_2$ have the same values as for \mbox{$T = 500$~K}.  \\

\noindent{\sc $R(\textrm{H})$:} 
For atomic hydrogen the de-excitation collisional rates are given for temperatures $T_{\textrm{kin}} \leq 2$\,$000$~K in the Leiden database using the data by \cite{Barinovs.etal.2005}. \cite{Goldsmith.etal.2012} fitted the data in the range \mbox{$20$~K~$\leq T_{\textrm{kin}} \leq 2$\,000~K} and found the relation between the coefficients and the kinetic temperature to be
\begin{equation}
\label{eq.H}
R_{\textrm{ul}}(\textrm{H}) = 7.6\times 10^{-10} \left(\frac{T_{\textrm{kin}}}{100 \textrm{\,K}}\right)^{0.14} \textrm{cm}^3 \textrm{s}^{-1}.
\end{equation}
We assume that the same relation holds at \mbox{$T > 2$\,000~K}. \\

\noindent{\sc $R(e^-)$:} 
De-excitation rate coefficients for collisions with electrons are given for \mbox{$T_{\textrm{kin}} \leq 20$\,$000$~K} in the Leiden database \citep{Wilson.Bell.2002}. For the coefficients at larger temperatures we use again a fit by \cite{Goldsmith.etal.2012}
\begin{equation}
\label{eq.e}
R_{\textrm{ul}}(e^-) = 8.7 \times 10^{-8} \left(\frac{T_{\textrm{e}}}{2000 \textrm{\,K}}\right)^{-0.37} \textrm{cm}^3 \textrm{s}^{-1},
\end{equation}
where we replace the electron temperature $T_{\textrm{e}}$ by the kinetic temperature $T_{\textrm{kin}}$.

\section{Testing Larson's microturbulence}
\label{app.Micro}
\begin{figure}
\centering
\includegraphics[width=80mm]{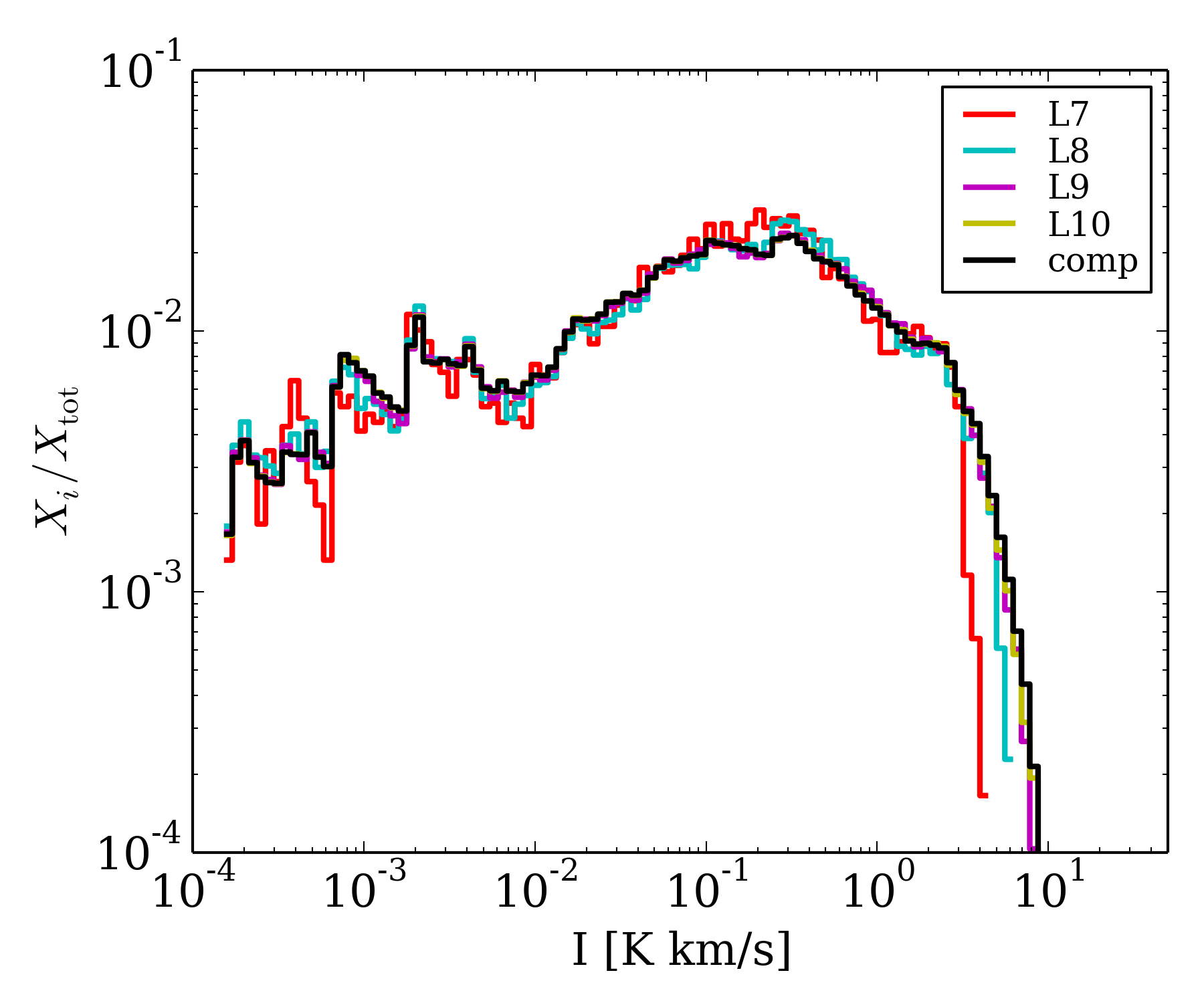}
\caption{Distribution of the intensities in the \CII\ emission maps (opacity affected) at different resolution, calculated assuming a microturbulence according to Larson~(1981). $X_i / X_{\textrm{tot}}$ denotes the fraction of pixels in every intensity bin. For comparison, the distribution of intensities for the simulation at L10, for which a thermal microturbulence was assumed, is plotted as black line. The microturbulence has only a negligible influence on the emission maps, and does not influence the behaviour of convergence among the resolution levels. }
\label{fig.TestMicro}
\end{figure}
To test whether our choice of the microturbulence influences our results, we additionally calculate the synthetic \CII\ line emission maps assuming a microturbulence according to \cite{Larson.1981}, following his eq.~(20) that is
\begin{equation}
\frac{a_{\textrm{turb}}}{\textrm{km}\,\textrm{s}^{-1}} = 1.1 \times \left(\frac{L}{\textrm{pc}}\right)^{0.38}.
\end{equation}
In Fig.~\ref{fig.TestMicro} we show the distribution of the intensities of those \CII\ line emission maps (opacity affected) for the different resolution levels with a bin size of $\Delta \textrm{log}_{10}(I\;[\textrm{K\,km\,s}^{-1}]) = 0.05$. For comparison, we plot the distribution for the simulation at L10, as calculated in Section~\ref{sec.Resolution} with the thermal microtubulence (black line). There are only minor differences between the maps at L10 with the different microturbulence, and no improvement in convergence when taking a microturbulence according to Larson.

\section{Testing escape probability lengths}
\label{app.EscProb}
\begin{figure}
\centering
\includegraphics[width=80mm]{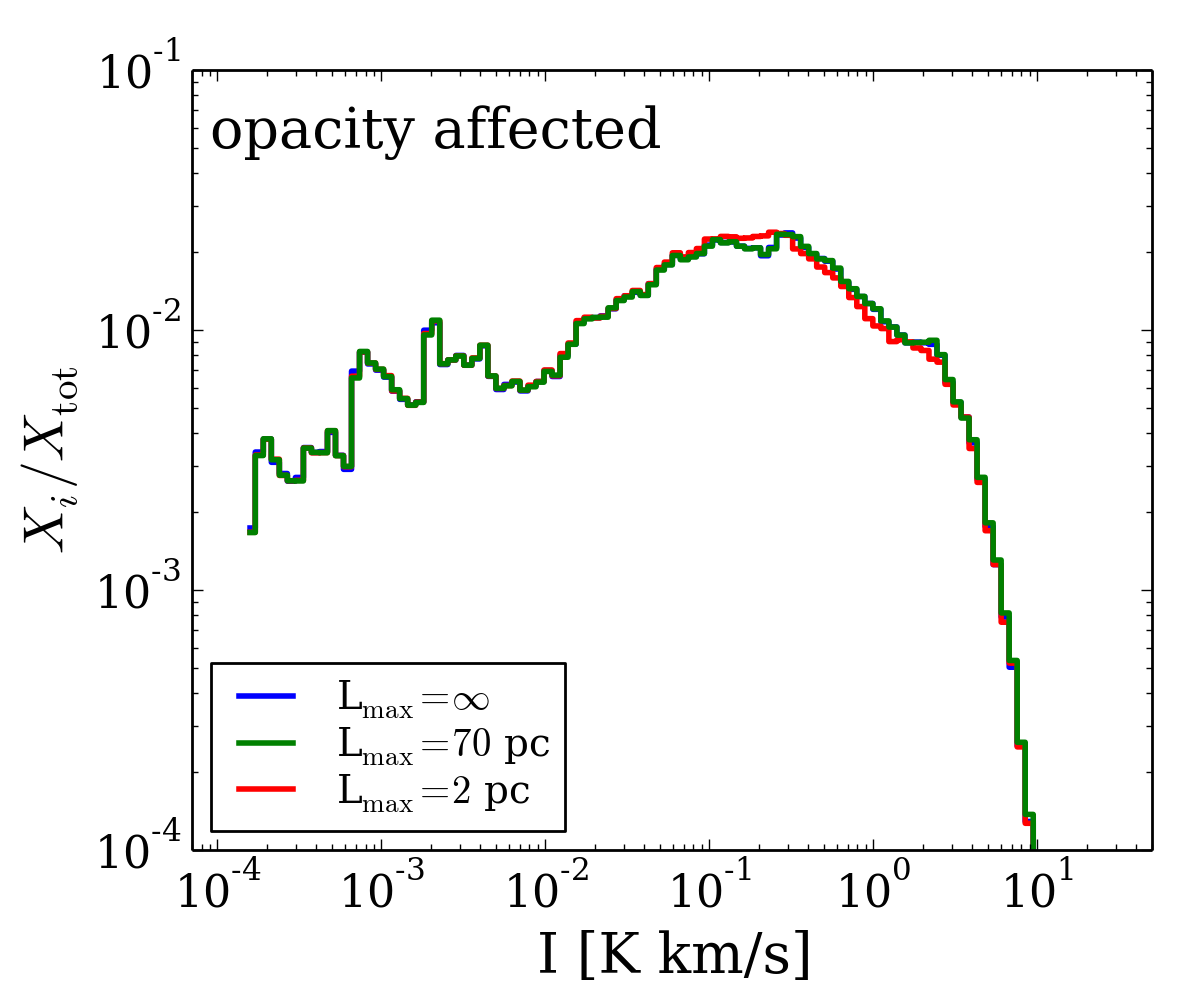}
\caption{Distribution of the intensities for the simulation at L10, when assuming different escape probability length scales. $X_i / X_{\textrm{tot}}$ denotes the fraction of pixels in every intensity bin. $L_{\textrm{max}} = \infty$ is chosen for the calculations within this work. Specifying $L_{\textrm{max}}$ for the simulation has only a negligible influence on the synthetic emission map.}
\label{fig.EscProb}
\end{figure}
\begin{figure}
\centering
\includegraphics[width=80mm]{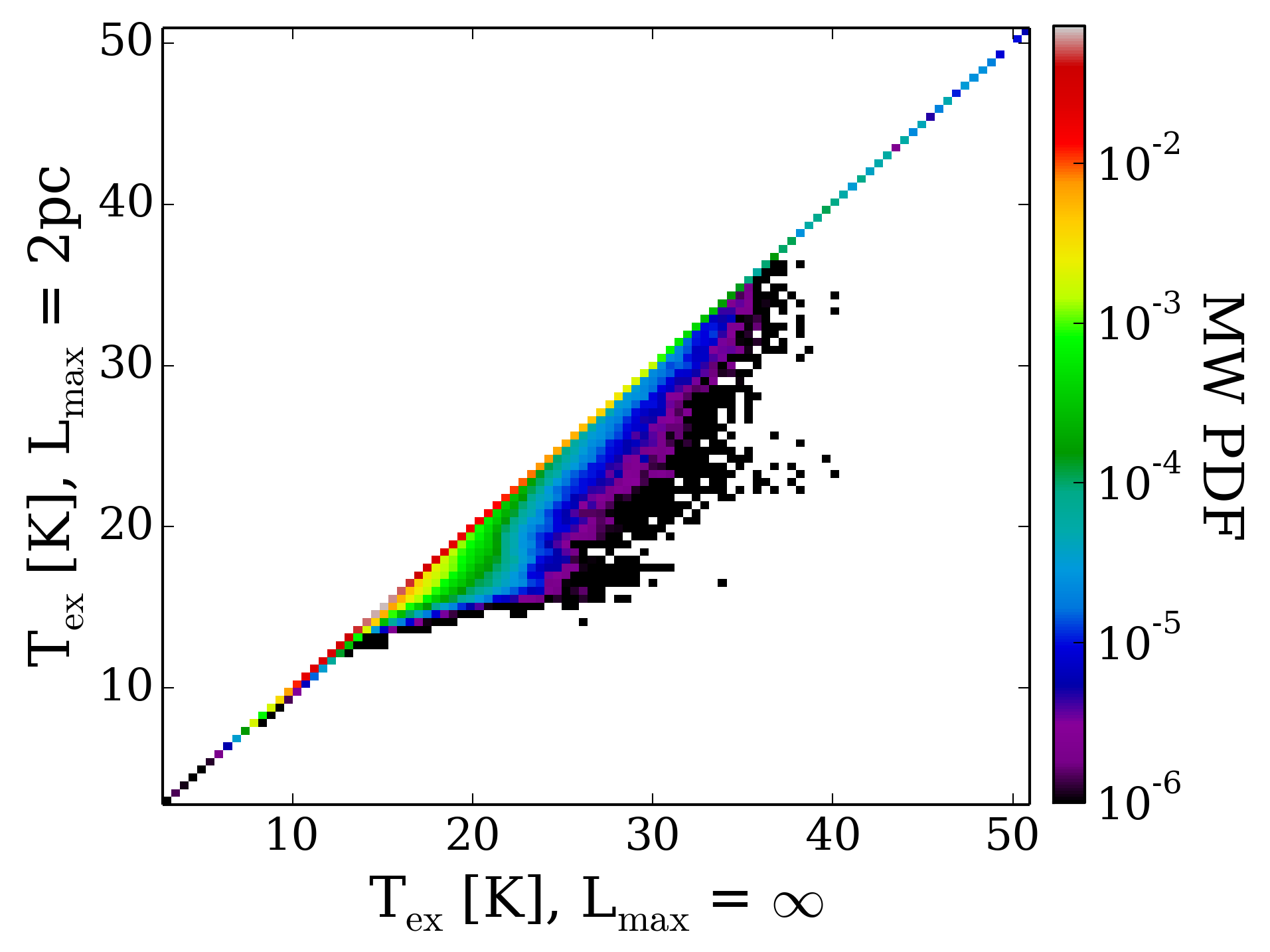}
\caption{Mass-weighted 2D-PDF of the excitation temperatures \Tex\ for the opacity affected \CII\ line emission calculated without an escape probability length ($L_{\textrm{max}} = \infty$, $x$-axis) and with $L_{\textrm{max}} = 2$~pc ($y$-axis). Colour-coded is the fraction of mass in the distribution. A similar plot comparing the calculation with optical depth effects and the optically thin case is shown in Fig.~\ref{fig.Tex_1213}. }
\label{fig.EscProb_Tex}
\end{figure}
When calculating the optical depth for the radiative transfer (Eq.~\eqref{eq.tau}), a velocity gradient is required. If the velocity gradient becomes too small, or a velocity field is not even given in the simulation, there is the possibility to specify an escape probability length scale to \RADMC, so that it calculates the optical depth as 
\begin{equation}
\tau_{ul}^{\textrm{L}} = \frac{ch}{4\pi} \frac{n_{\textrm{C}^+} L_{\textrm{max}}}{\sqrt{\pi}  a_{\textrm{tot}}} \left(x_l B_{lu} - x_u B_{ul} \right).
\end{equation}
Typically, $L_{\textrm{max}}$ is set to the size of the simulation box. If an escape length probability is given, \RADMC\ calculates $\tau_{ul}^{\textrm{LVG}}$ and $\tau_{ul}^{\textrm{L}}$ and takes the minimum of it for the radiative transfer calculation. We test whether setting an escape probability length scale influences the resulting synthetic emission maps.  Figure~\ref{fig.EscProb} shows the distribution of the intensities when no escape probability length is given ($L_{\textrm{max}} \rightarrow \infty$), and when it is set to 70~pc and 2~pc. An $L_{\textrm{max}}$ of 70~pc corresponds to the size of the simulation box, and $L_{\textrm{max}} = 2$~pc we chose for reason of comparison ($\Delta \textrm{log}_{10}(I\;[\textrm{K\,km\,s}^{-1}]) = 0.05$). As can be seen in Fig.~\ref{fig.EscProb}, the escape probability length scale has only an negligible influence on the intensity distribution. Figure~\ref{fig.EscProb_Tex} shows 2D-PDF (histogram) of the scatter plot between the excitation temperatures calculated with $L_{\textrm{max}} = \infty$ and $L_{\textrm{max}} = 2$~pc. This plot is similar to the one shown in Fig.~\ref{fig.Tex_1213}, where the excitation temperatures of the \CII\ line emission in the opacity affected and the optically thin case are presented. The $x$-axes in both plots are identical. When setting $L_{\textrm{max}} = 2$~pc, \Tex\ is similar to \Tex\ of the optically thin emission for low intensities (cf. Fig.~\ref{fig.Tex_1213}). For higher intensities, the excitation temperature of the calculation with $L_{\textrm{max}} = 2$~pc has intermediate values between the optically thick and thin case. The smaller the value of $L_{\textrm{max}}$ is chosen, the more \Tex\ approaches to the optically thin solution.

\section{Spectral resolution}
\label{app.SR}
\begin{figure}
\centering
\includegraphics[width=80mm]{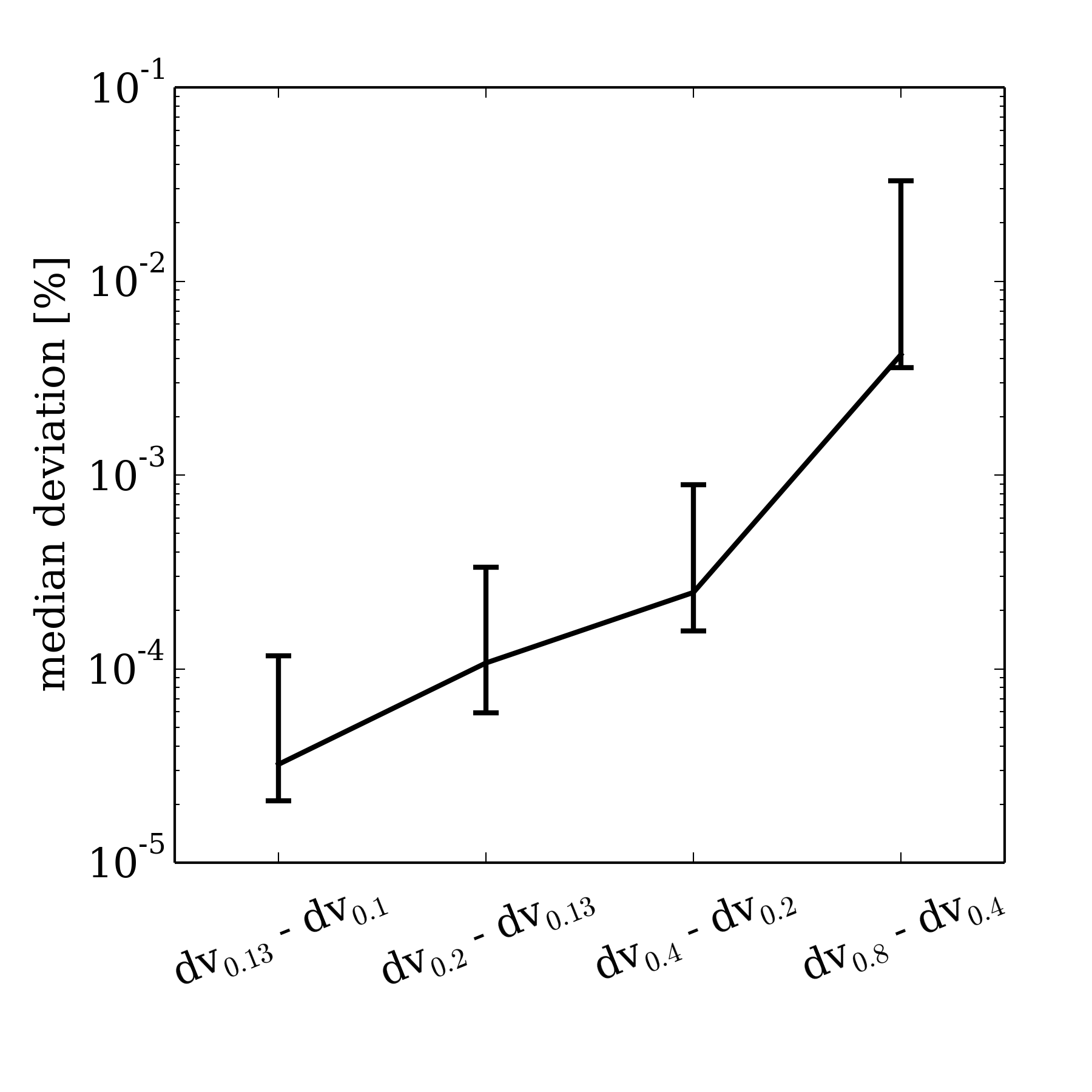}
\caption{Deviation of the synthetic \CII\ line emission maps (opacity affected) at different spectral resolutions.}
\label{fig.Mom_channels}
\end{figure}
In the synthetic \CII\ emission maps for Section~\ref{sec.Resolution} we chose a spectral resolution of d$v = 0.2$~\kms, inspired by the spectral resolution of the integrated line maps of the \textit{Herschel} \CII\ line emission in Orion. As the velocity range of $\pm 20$~\kms\ in our synthetic observations is fixed, this corresponds to 201 channel maps. We test different spectral resolutions by setting the amount of channels to 401, 301, 201, 101 and 51, corresponding to d$v = 0.1, 0.13, 0.2, 0.4, 0.8$~\kms, respectively. We calculate for every pixel $p$ in two spectral adjacent emission maps ($p^{\textrm{d}v_x}$, $p^{\textrm{d}v_{(x+1)}}$, respectively) the deviation of the integrated intensities in percentage by 
\begin{equation}
\Delta_v = (p^{\textrm{d}v_x} - p^{\textrm{d}v_{(x+1)}}) \times \frac{100\%}{p^{\textrm{d}v_x}}.
\end{equation}
We derive the median and the interquartile range of the absolute value of the distribution of $\Delta_v$. The median is the value for that 50\% of the distribution is included, and the interquartile range the difference at which 25\% and 75\% of the distribution are included. We show the result in Fig.~\ref{fig.Mom_channels}, where we present on the $y$-axis the median of the deviation $\Delta_v$ with the interquartile range as error bars, and indicate on the $x$-axis the spectral resolutions of the synthetic emission maps. Since the median deviation of $\Delta_v$ is less than $0.001$\% between the spectral resolutions d$v = 0.1$ and d$v = 0.2$~\kms, we conclude that a spectral resolution of d$v = 0.2$~\kms\ is sufficient to capture the \CII\ line emission. \\

\section{Convergence studies}
\subsection{Convergence of the emission maps}
\label{app.ConvMaps}

\begin{figure*}
\centering
\includegraphics[width=\textwidth]{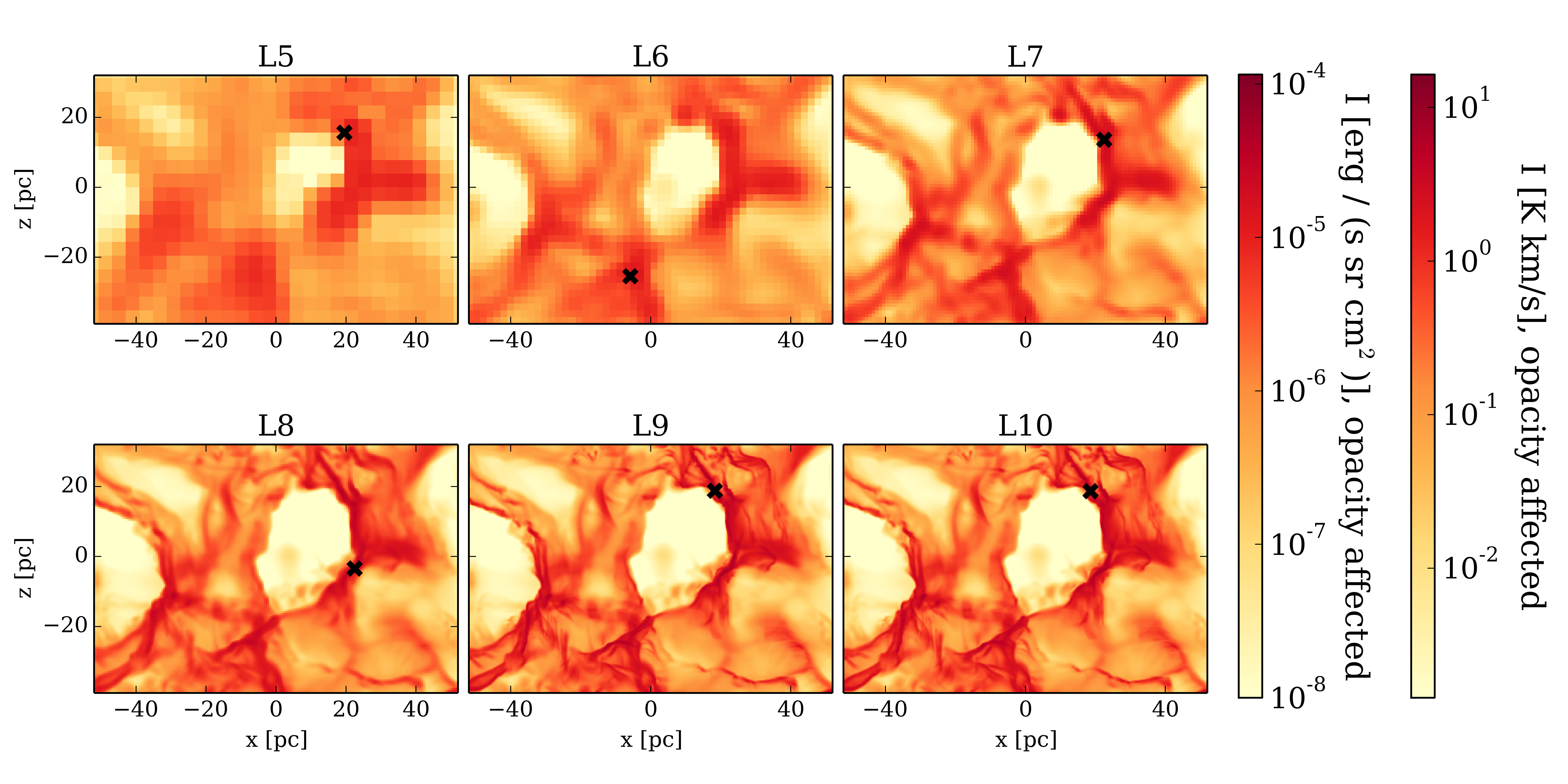}
\caption{Synthetic \CII\ line emission maps (opacity affected) of the zoom-in simulation at different resolution levels as listed in Table~\ref{tab.ZoomIn}. The crosses mark the position with the maximum integrated intensity in each map. From the refinement level L5 to L9 this position changes due to varying optical depths. }
\label{fig.Mom0_Res}
\end{figure*}
\begin{figure*}
\centering
\includegraphics[width=\textwidth]{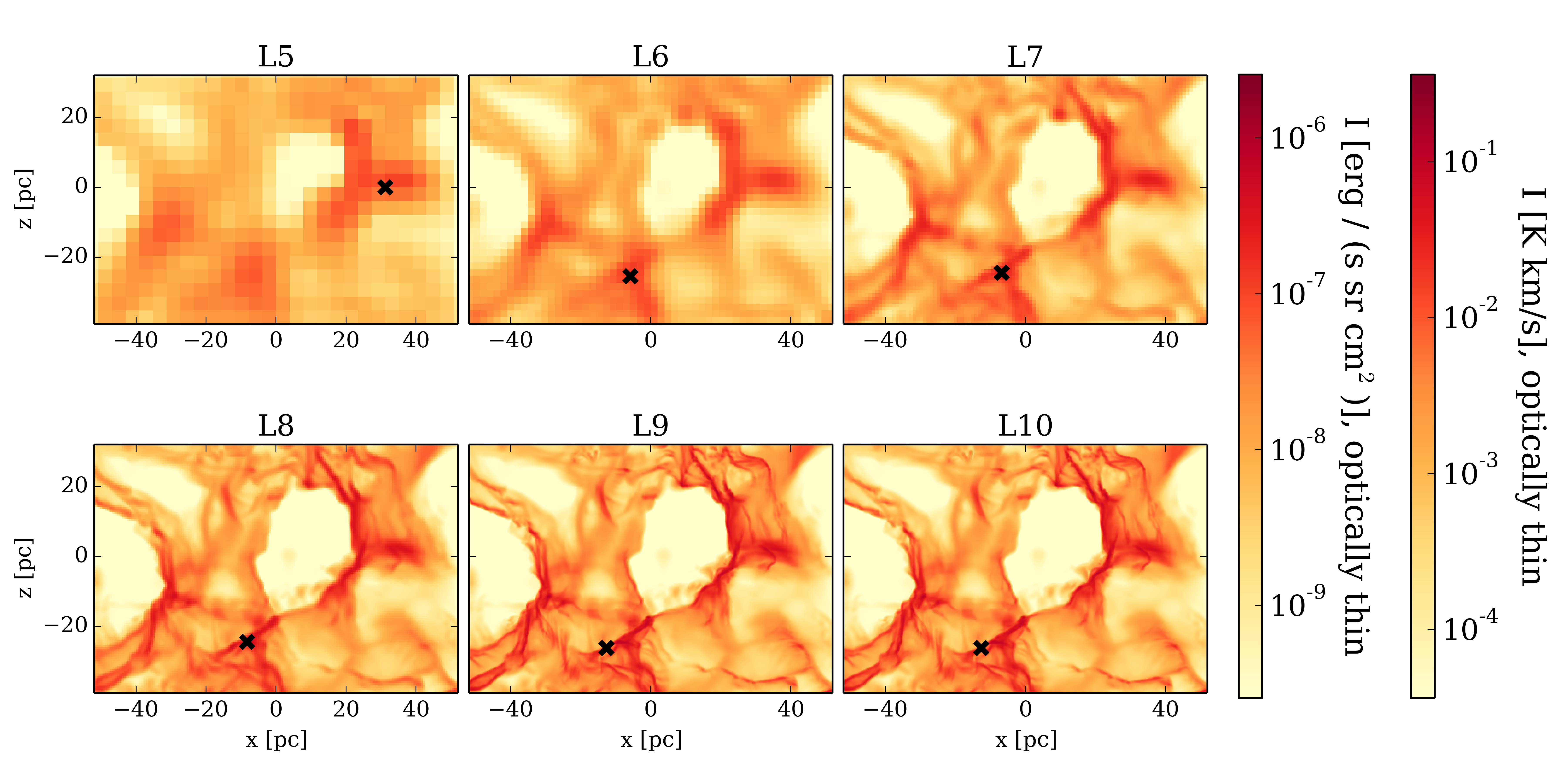}
\caption{Similar to Fig.~\ref{fig.Mom0_Res}, the synthetic \CII\ line emission maps for the optically thin case. The crosses mark the position with the maximum integrated intensity in each map. Here, the position remains in the same area from L6 on, as the emission is optically thin. }
\label{fig.Mom0_Res_CII13}
\end{figure*}
We investigate whether and how the change in resolution influences the synthetic \CII\ line emission maps in the opacity affected and optically thin case. Figs.~\ref{fig.Mom0_Res} and \ref{fig.Mom0_Res_CII13} show the \CII\ synthetic emission maps for these cases for all resolution levels for the \textit{x-z} projection at an evolutionary time of $t_{\textrm{tot}} = 13.9$~Myr. The crosses mark the positions of the maximum integrated intensity for each resolution. The values of the peak intensities and the luminosities integrated over these  maps are summarised in Table~\ref{tab.ZoomIn} in the fourth to seventh column. For opacity affected \CII, the position of the maximum of the integrated intensity jumps around for different refinement levels (L5 to L9), although the structures within the molecular clouds maintain their general morphology for all resolution levels. Regions that were optically thin before can become optically thick with a higher resolution, because the density fluctuations along one line of sight become more distinct. For the optically thin \CII\ synthetic emission maps (Fig.~\ref{fig.Mom0_Res_CII13}) the position of the maximum integrated intensity changes only a little from refinement level L6 on, since the emission is not influenced by optical depth effects. \\

\begin{figure*}
\centering
\includegraphics[width=\textwidth]{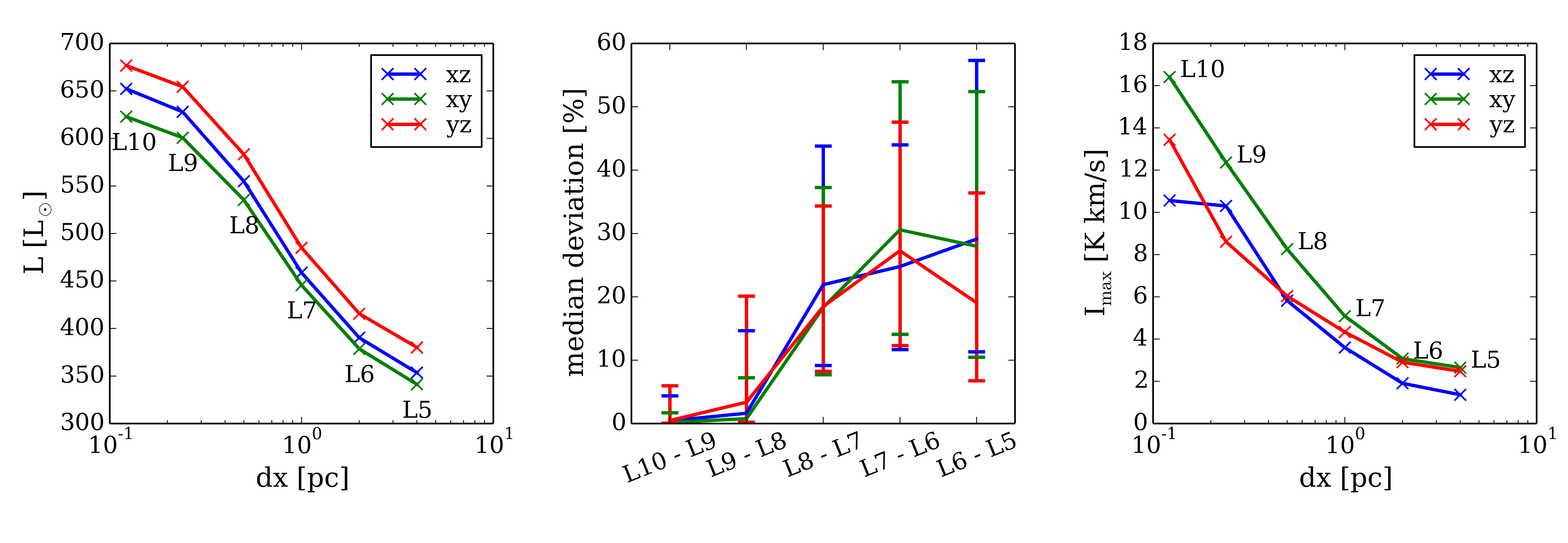}
\caption{Total \CII\ luminosity (opacity affected; left) for different resolution levels as listed in Table~\ref{tab.ZoomIn}, shown for all three projections. In the middle panel the median of the deviation between the synthetic emission maps of two adjacent resolution levels is shown. As error bars we use the interquartile range. The median of the deviation between the map of L9 and L10 is $< 0.5$\%. In the right panel, the peak intensities against the resolution level is plotted, which is not converged. }
\label{fig.ZoomIn_Peak}
\end{figure*}
\begin{figure*}
\centering
\includegraphics[width=\textwidth]{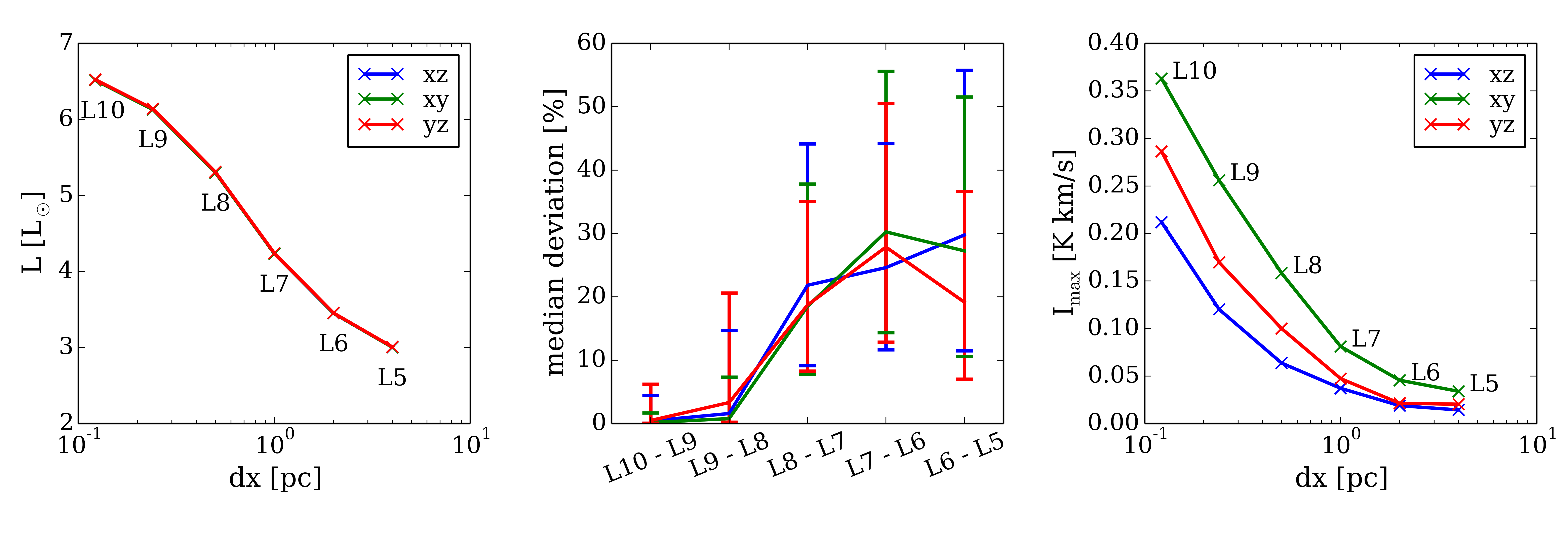}
\caption{The same as Fig.~\ref{fig.ZoomIn_Peak}, but for the synthetic \CII\ line emission maps in the optically thin case. For the total luminosity all projections fall on to one curve. }
\label{fig.ZoomIn_Peak_CII13}
\end{figure*}
Figs.~\ref{fig.ZoomIn_Peak} and \ref{fig.ZoomIn_Peak_CII13} show in the left panels the \CII\ luminosities integrated over the whole maps for the opacity affected (Fig.~\ref{fig.ZoomIn_Peak}) and optically thin case (Fig.~\ref{fig.ZoomIn_Peak_CII13}) as a function of the maximum refinement level. The best refinement level L10 with d$x = 0.122$~pc is on the left hand side of each plot. For all three projections in both cases the luminosity increases with higher spatial resolution. Note that for the optically thin synthetic \CII\ line emission the luminosities are the same for each projection. The difference between the maps of two adjacent resolution levels decreases with smaller d$x$. This can be seen in the middle panels, where we compare the integrated intensity values of the maps of two adjacent resolution levels, L$x$ and L$(x+1)$. First we reduce the amount of pixels of the higher resolution to the lower one by averaging the values of four pixels $p^{x + 1}_i$ in each map of the resolution level L$(x+1)$ to one value. Then we compare the result with the pixel value $p^{x}$ of the map of level L$x$:
\begin{equation}
\Delta = \left(p^x - \frac{1}{4} \sum_{i = 1}^{4} p^{x + 1}_i \right) \times \frac{100\%}{p^x}. 
\end{equation}
We take the median of the absolute values of the differences $\Delta$ for all pixels and present them on the $y$-axis of the plot. The median is the point at which 50\% of the distribution is included. The $x$-axis indicates the corresponding resolution levels. The error bars are given by the interquartile range of the distribution of $\Delta$, which are those points at which 25\% and 75\% of the pixels in the distribution are included. This technique has the advantage of being robust against individual spikes in a distribution. As seen from this plot the deviation decreases for all projections when going to higher resolution levels. This is in agreement with the smaller increase of the luminosity of the maps with higher resolution level, as shown in the left panels of Figs.~\ref{fig.ZoomIn_Peak} and \ref{fig.ZoomIn_Peak_CII13}. The intensity distributions of the maps converge within 0.5\% at L9 to L10. The right panels shows the peak intensity for every projection as a function of d$x$. Since the position of the maximum integrated intensity is not fixed and the volume density still increases with higher resolution level, we do not expect convergence for \CII\ in this property. \\

\subsection{Chemical composition at different resolution levels}
\label{app.L6L10}

\begin{figure}
\centering
\includegraphics[width=80mm]{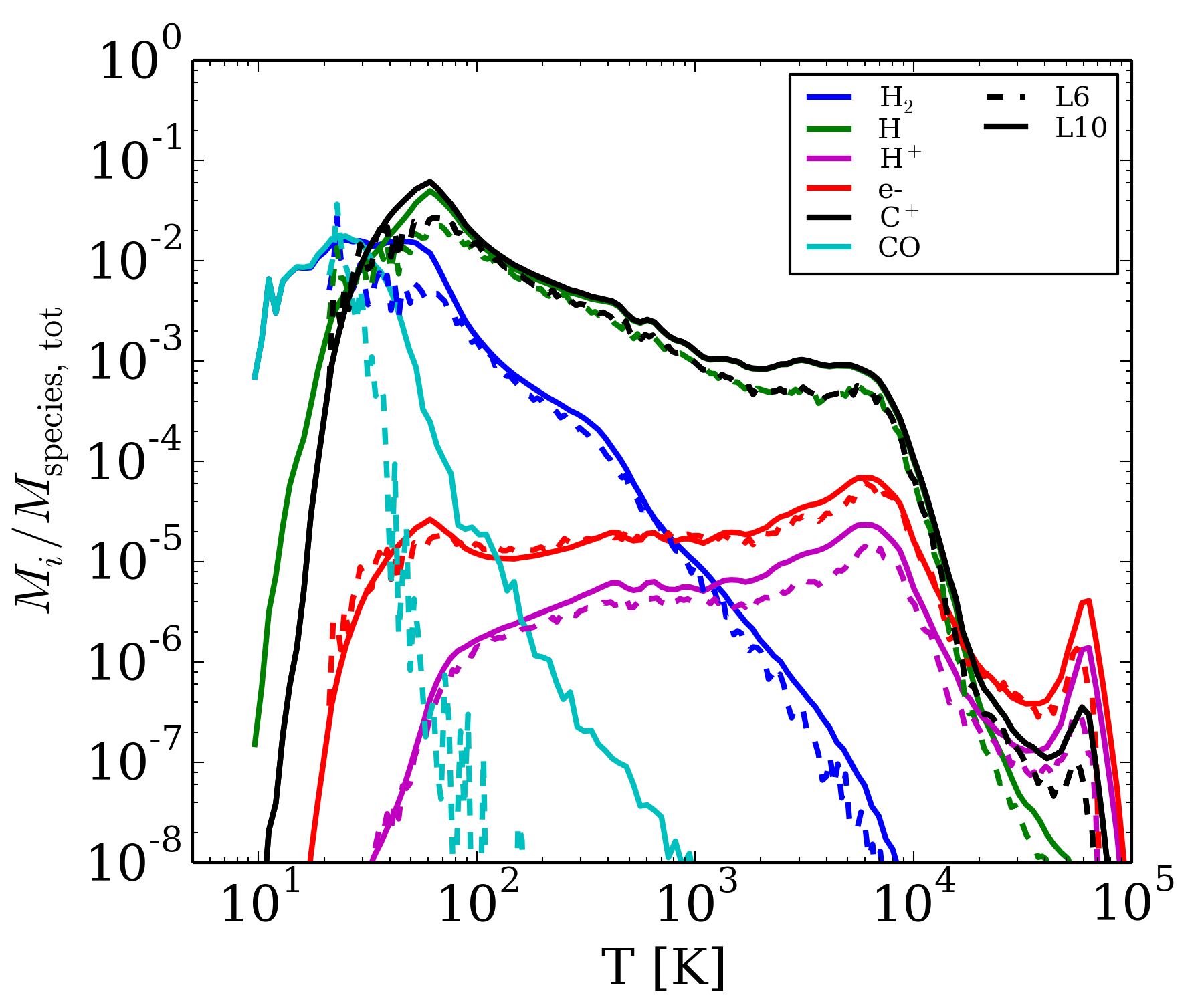}
\caption{Comparison of the mass-weighted temperature distributions for all species for the resolution level L6 (dashed line) and the highest resolution level L10 (solid line). $M_i / M_{\textrm{species,~tot}}$ denotes the mass fraction of each species normalized to the total hydrogen mass (for H and H$_2$), the total carbon mass (for C$^+$ and CO), or the total electron mass, respectively (see Section~\ref{sec.Analysis_Temp}). }
\label{fig.ZoomIn_Origin_T_comp}
\end{figure}
To see whether the abundances of the collisional partners as a function of the temperature change with the resolution level, we show the mass-weighted temperature distributions of all chemical species for the levels L6 (dashed lines) and L10 (solid lines) in Fig.~\ref{fig.ZoomIn_Origin_T_comp}. In general, the gas at low temperatures is better resolved for L10, down to \mbox{$T \sim 10$~K} for all chemical species. For L6, the gas has only temperatures above 20~K and CO is not resolved. For the other species, the qualitative behaviour of the distributions are similar for $T \gtrsim 100$~K. Between 50~K and 100~K, H$_2$, H and C$^+$ are less abundant in L6 compared to L10. This explains the lower luminosity we find in L5 (cf. Figs.~\ref{fig.ZoomIn_Peak} and \ref{fig.ZoomIn_Peak_CII13} and Table~\ref{tab.ZoomIn}). In general, the distributions for H and C$^+$ are aligned for the L6 and L10 simulations. \\

\bibliographystyle{mn2e}
\bibliography{Literature_AF}

\label{lastpage}

\end{document}